%% file: ms.tex
\documentclass[journal=nalefd,manuscript=article,layout=traditional]{achemso}

\usepackage[version=3]{mhchem} 

\usepackage{graphicx}
\usepackage{dcolumn}
\usepackage{bm}
\usepackage{xcolor}
\usepackage{multirow}
\usepackage{makecell}
\usepackage{amsfonts}
\usepackage{placeins}
\usepackage{soul}
\usepackage{hyperref}
\author{Jakub \v{S}ebesta}
\email{jakub.sebesta@physics.uu.se}
\author{Oscar Gr\r{a}n\"as}
\affiliation[UU]{Materials Theory, Department of Physics and Astronomy, Uppsala University Box 516, 751 20 Uppsala, Sweden}
\email{oscar.granas@physics.uu.se}

\title{Photo-induced modification and relaxation dynamics of Weyl-semimetals}

\begin{document}

%
%
%
%
%

\begin{abstract}
  The use of ultrashort laser pulses to investigate the response of materials on femtosecond time-scales enables detailed tracking of charge, spin and lattice degrees of freedom. When pushing the limits of the experimental resolution, connection to theoretical modeling becomes increasingly important in order to infer causality relations. Weyl-semimetals is particular class of materials of recent focus due to the topological protection of the Weyl-nodes, resulting in a number of fundamentally interesting phenomena. In this work, we provide a first-principles framework based on time-dependent density-functional theory for tracking the distribution of Weyl-nodes in the Brillouin-zone following an excitation by a laser pulse. For the material TaAs, we show that residual shifts in the Weyl-Nodes’ position and energy distribution is induced by a photo-excitation within femto-seconds, even when the laser-frequency is off-resonant with the Weyl-node. Further, we provide information about the relaxation pathway of the photoexcited bands through lattice vibrations. 
\end{abstract}

\input{nanolett_intro_V1}

\input{nanolett_results_V1} 

\input{nanolett_conclusions}

\begin{acknowledgement}
The Carl Trygger foundations is acknowledged for funding through grant CTS20:153. O.G. Acknowledges financial support from the Swedish Research Council (VR) through grant 829 2019-03901 and European Research Council through the Synergy Grant 830 854843 - FASTCORR. The computations were/was enabled by resources provided by the National Academic Infrastructure for Supercomputing in Sweden (NAISS) and the Swedish National Infrastructure for Computing (SNIC) at NSC and PSC, partially funded by the Swedish Research Council through grant agreements no. 2022-06725 and no. 2018-05973.
\end{acknowledgement}

%
%

\section{Supporting Information}

\input{nanolett_supportinfo}
\bibliography{apssamp,weyl}
\end{document}

%% file: nanolett_intro_V1.tex
\section{Introduction}

Materials where the electronic band-structure exhibits non-trivial topological states has garnered significant interest recent days. In particular, semi-metals where the band-structure forms so-called Weyl-nodes (WN), 
may show outstanding physical phenomena such as negative magnetoresistance, anomalous Hall effect, non-local transport or quantum oscillations in the magnetotransport~\cite{r18_Armitage_WS_DS_rev,r17_Felser_WSrev,r17_Wang_WSM_tranport_rev}. The topological properties has also been argued to provide an energy efficient avenue for information storage and manipulation \cite{Sie:2019cf}

The topological protection is manifested by the presence of pairs of topologically protected band crossings occurring nearby high symmetrical lines in the bulk material~\cite{r18_Armitage_WS_DS_rev,r17_Felser_WSrev,r17_Wang_WSM_tranport_rev,r17_IFF48_LNotes,r11_Burkov_WSM_prediction,r07_Murakami_3Dgapless}. Their presence in the ground-state does not require any special symmetry protection other than the crystal symmetry, making the points  stable under any adiabatic local perturbation. 
The band crossings form Weyl cones touching at the WNs~\cite{r15_Xu_WeylF,r15_Lv_TaAs}. Each pair hosts Weyl quasiparticles with different chirality at the coupled WNs. They represent vortices of the Berry phase~\cite{r10_Di_Bphase_rev,r17_IFF48_LNotes}, namely the monopole and antimonopole
of the Berry curvature~\cite{r18_Armitage_WS_DS_rev,r17_Felser_WSrev,r17_Wang_WSM_tranport_rev,r21_Xie_TopSemM_rev}. It is characterized by a non-vanishing topological invariant so-called Chern number $C$~\cite{r18_Armitage_WS_DS_rev}. At WNs, it  acquires non-zero values depending on the vortex character. 


The stability of the WNs is tightly connected to the source- and drain properties of the Berry-curvature, and an annihilation  may only occur by merging Weyl-points of opposite Chern numbers.
Therefore, the manipulations of topological states are in practice often performed through modulation of lattice degrees of freedom through means of Thz radiation \cite{Sie:2019cf,Xiao:2020gk}, optical pumping \cite{Ji:2021cn,ShengMeng:2021} or combination of nano-structuring and an external pump \cite{ShaozhengJi:2022gb}. 
From a modeling perspective, much conceptual work was performed on low-energy models \cite{r18_Armitage_WS_DS_rev}. However, under strong pumping the quasi-particle band-structure may change significantly \cite{Granas:2022eu}, and the validity of low energy models is challenged. Therefore, for direct comparison with experimental data in pump-probe situations, a description of the full band-structure is needed.
In previous work, Shin \emph{et al.} used the velocity field to track the anomalous features of the conductivity \cite{Shin:2019kt}, employing the real-time version of time-dependent density functional theory (RT-TDDFT) in a pseudo-potential framework \cite{Octopus_code}.


If one was able to extract the  time dependent quasi-particle band structure, the introduced scheme could be employ to follow a WN dynamics induced \emph{e.g.} by laser pulse. 
The solution could be the TD-DFT approach~\cite{Sharma2014_TDDFT,r02_Onida_TDDFT_rev,r99_Kohn_NLect,r84_RungeGross_TDDFT}. Mostly, it is used to determine time dependent  integral quantities   \emph{e.g.} the density of states (DOS) or magnetization.
 Nevertheless, there arise a few simple approaches to obtain the time evolved band structure as well. 

In this work, we implement a theoretical framework that allows us to investigate the impact of strong electromagnetic fields on the WN dynamics in a materials realistic setting. We separate the impact of strong fields and relaxation of the pumped state through lattice motion, and 
{provide a framework for investigating  response functions in order to deduce how the general susceptibility changes in the material}. Our developments are focused on the all-electron implementation of RT-TDDFT in the \emph{Elk} code \cite{ElkCODE}. 

Particularly, we employ our methods to study the impact of a laser pulse  on the well-know compound TaAs, a simple prototype of Weyl semimetals, where its non-trivial  band structure topology has a crucial impact on its physical properties. TaAs crystalize in the I4$_1$md space group missing the xy-mirror plane  As a result of the crystal inversion symmetry broken (characteristic for nonmagnetic Weyl materials) which give rise robust so-called Weyl cones with linear dispersion occuring in  the band structure, related to  hosting of chiral massless Weyl quasiparticle states.

%% file: nanolett_results_V1.tex
\section{Results } 

%
%
%
%


\subsection{Ground state}

Before the TD-DFT treatment, the ground state calculations without an applied external laser field were performed.  They especially served to demonstrate the ability to seek WNs and determine WNs' position within the BZ to facilitate the evaluation of the TD-DFT calculations. The obtained ground state electronic band structure (Fig.~\ref{Fig:band_reco}) corresponds well to the literature~\cite{r15_Xu_WeylF}. For the selected k-path, the valence and conduction bands are almost touching near the Fermi level $E_{F}$,  between the $\Sigma_{\mathrm{1}}$, N and $\Sigma$ points, likely indicating a presence of WNs.  Except for these region, the conduction and valence bands stay apart which  corresponds to a semi-metal character of the TaAs~\cite{r15_Lv_TaAs}.

Having evaluated the band structure, we localized the presence of WNs. For a fast and effective searching  WNs' positions, we initially divided the BZ into several slabs to roughly determine the WNs' position (Eq.~\ref{Eq.Chern_gamma}). Later, by squeezing the   size (Eq.~\ref{Eq:gamma_Usegments}) of the {Wilson loop} \cite{r11_Yu_z2invariant_berry_connection}, we traced the WNs more accurately.


\begin{figure*}[t]
\includegraphics[width=\textwidth]{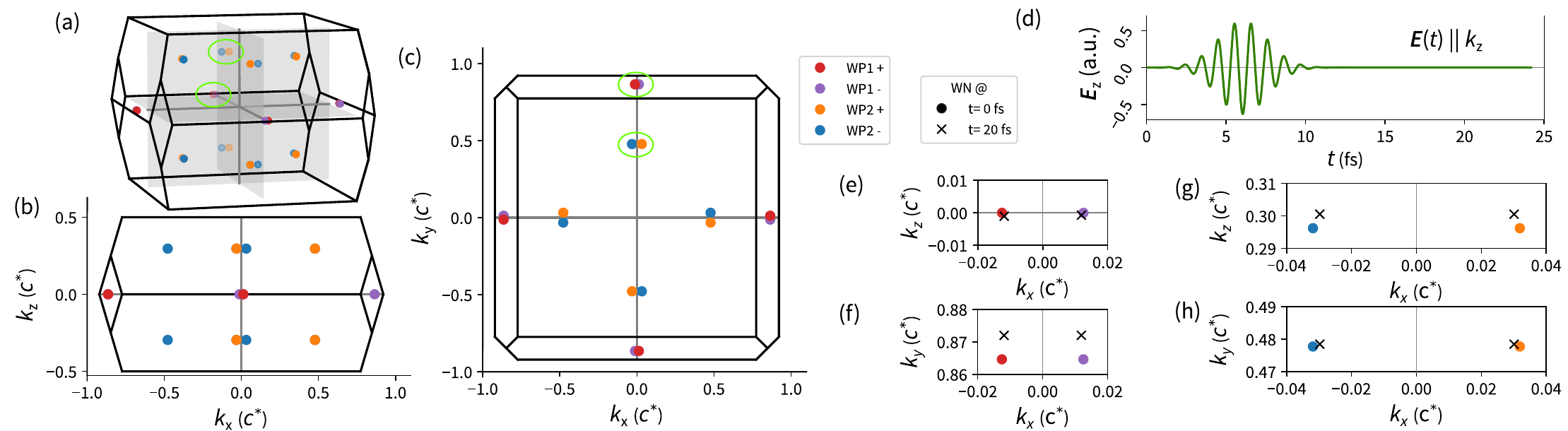}
\caption{Relative Weyl nodes shift induced by laser pulse. (a-c) Position of Weyl nodes within the BZ at the ground state. (d) Applied effective electric field. Laser power 3.9~eV and laser fluence 10 mJ$/$cm$^{2}$. Field was parallel to the $k_{\rm{z}}$ axis. (e-h) Shift of the Weyl nodes after the laser pulse. Weyl nodes' positions at (bullet) $t$=0~fs and (cross) $t$=17~fs are compared.}
\label{Fig:WN_overview}
\end{figure*}

It is worth mentioning that the determined WN positions are quite strongly affected by the loop size, which is expressed by shifting the position of the discontinuity in the integrated phase along a studied direction in the k-space (Fig.~\ref{Fig:WN_loopsize_effects}). However, a convergence with respect to the k-spacing can be achieved. Besides, for a too-large loop, a not well-separated WN pair might be hidden as the difference is integrated out 



Corresponding to the literature, we observed two sets of WNs (Fig.~\ref{Fig:WN_overview}). First, 4 pairs of W1 WNs lying in the $k_{\mathrm{x}}k_{\mathrm{y}}$-plane were detected.
Regarding the calculated band structure (Fig.~\ref{Fig:band_reco}), they can be ascribed to almost touching bands at the Fermi level $E_{\rm{F}}$  near the $\Sigma$  point. 
Second,  8 pairs of W2 WNs possessing non-zero $k_{z}$ component. They are indicated by band proximity near the $\Sigma_{1}$ point (Fig.~\ref{Fig:band_reco}). WN mutual chiralities (Fig.~\ref{Fig:WN_overview}) are depicted based on the relative topological charge resulting from the sign of the integrated phase singularity (Fig.~\ref{Fig:WN_loopsize_effects}). W1 and W2 nodes differ not only by their $k_{\rm{z}}$ component but also by the WN pairs separation in the $k_{\mathrm{x}}k_{\mathrm{y}}$-plane. The nodes in the W1 pairs are about two times closer than W2 ones (Fig.~\ref{Fig:WP_pair_bands}) \cite{r15_Lv_TaAs,r15_Yan_noncentrosymetric_WS}. Besides, the W1 nodes lie about 14~meV below the W2 ones in the energy in agreement with the literature~\cite{r15_Xu_WeylF,r15_Yan_noncentrosymetric_WS}.
It suggests that the W2 nodes are likely more important for the magneto-transport chiral anomalies~\cite{r15_Yan_noncentrosymetric_WS}.


\subsection{Time Evolution}

Having verified the ground state  WNs' positions, we focused on the impact of an ultrafast laser pulse on WNs' behavior. Particularly, we were interested in their presence and modification with the respect to the ground state as a function of the increasing time delay.



To study a possible relaxation process after the laser pulse duration, a short laser pulse  width  FWHM$\sim${3.6~fs} was selected due to the numerical stability and computational demands. 
To obtain a quantitative scaling of the system response, two distinct pulse strength were considered (see~Supporting Information) 
%
%
%
%
%
%
%
However, for all the studied cases, a dismantling of the WNs by the laser pulse was not observed.
The presence of WNs was detected (Eq.~\ref{Eq.Chern_gamma}) (Figs.~\ref{Fig:TaAs_WP1}) for the reachable time range. 




\subsubsection{Band structure}

As expected, the applied laser pulse brought about modifications in bands' occupancies by electron state excitations (Fig.~\ref{Fig:Occ_strong_p}) as well as reconstruction of the bands itself (Fig.~\ref{Fig:band_reco}). Naturally, according to the used laser pulse, we observed different excitation from the valence band to the conduction one concerning the laser pulse energy. Along the considered k-path, the weaker laser pulse (P$_{\mathrm{B}}$) induced transition only at a few quite well localized hot spots between the band in the vicinity of the ground state Fermi level $E_{\mathrm{F}}^{\mathrm{0}}$ (Fig.~\ref{Fig:Occ_strong_p}d). Elsewhere, the change of occupancy is rather negligible or it doesn't change at all. We refer the energy to the ground state Fermi level $E_{\mathrm{F}}^{\mathrm{0}}$ in the text as all the occupations originate from the projection of the time evolved states to the initial ground state electronic states.
%
On the other hand, the stronger pulse (P$_{\mathrm{A}}$) gave rise to excitation into slightly higher conduction bands thanks to the higher laser pulse energy, where the modification on the occupancy spreads nearly across the entire studied k-path (Fig.~\ref{Fig:Occ_strong_p}b). In both cases (Fig.~\ref{Fig:Occ_strong_p}), similar conduction bands are depleted. However, for the stronger laser pulse (Fig.~\ref{Fig:Occ_strong_p}d), more possible transitions occur as the higher conduction bands are flatter.  Besides, a larger delivered amount of energy, expressed by a higher laser fluence, is attributed to the pulse P$_{\mathrm{A}}$. Nevertheless, the occupation nearby the WNs changed only negligibly (Fig.~\ref{Fig:Occ_strong_p}).

Along the occupation modifications, a reconstruction of the band structure took place (Fig.~\ref{Fig:band_reco}), where the time dependent bands are related to the eigenvalue spectra of the Houston states~(Eq.~\ref{Eq.:Houston_states}) ~\cite{r15_Wu_HoustonSt,r86_Krieger_HoustonBasis}. In general, a non-uniform shift, depending on the k-position, of the electronic bands towards higher energy was observed.  The effect is the more pronounced the higher the laser fluence is as the TaAs system absorbs a larger amount of energy. However, no intense bending of the bands was observed.
%

%
%

\subsubsection{Weyl nodes}

\begin{figure*}[t]
\includegraphics[width=\textwidth]{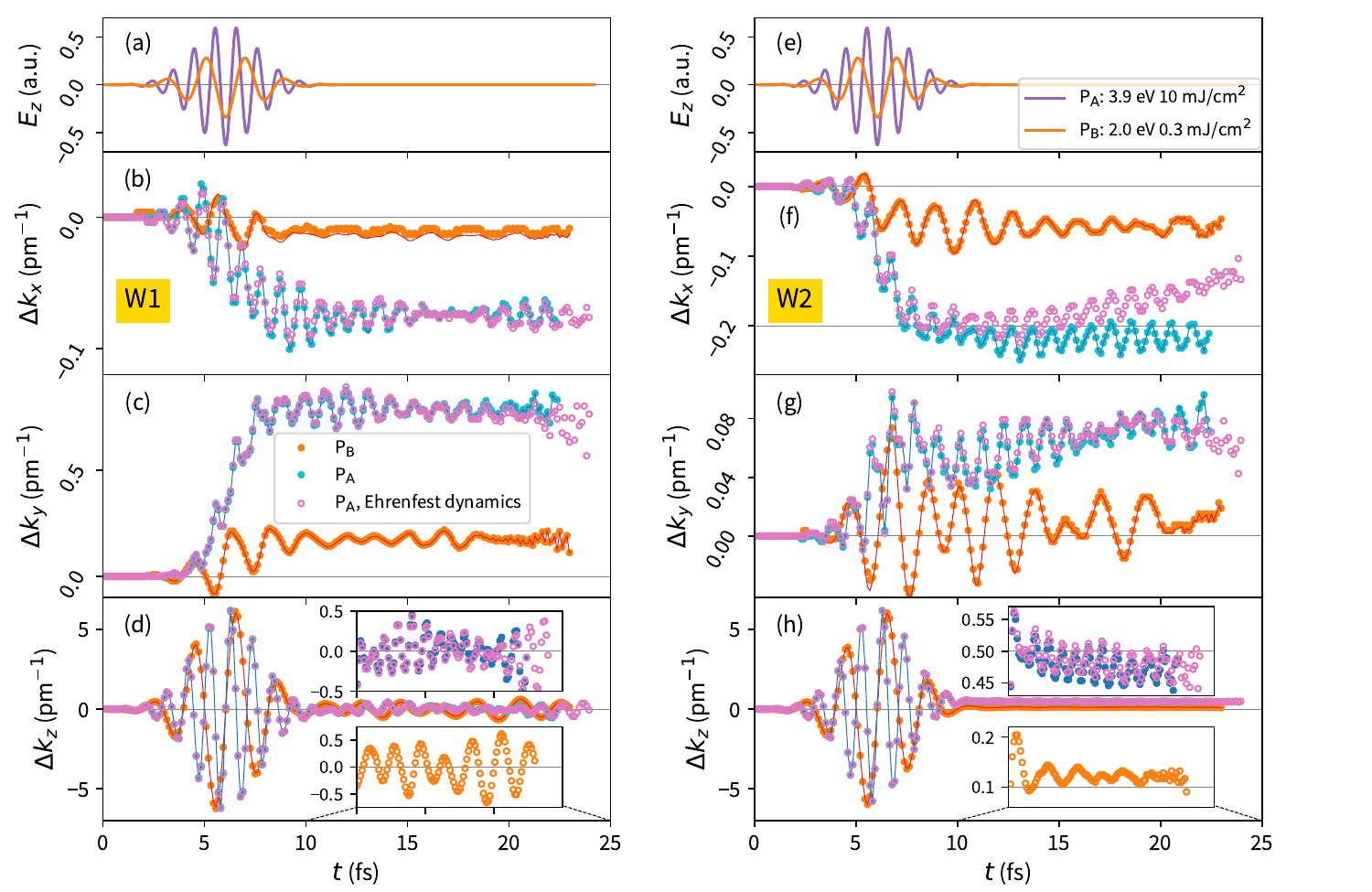}
\caption{TaAs Weyl-nodes' position time evolution. (left column) W1 node . (right column) W2 node.  (points) WN positions from Berry phase integration (lines) and from band structure are depicted. (filled points and lines) No ion dynamics included. (empty points) Ehrenfest dynamic included. Two different pulse strengths (P$_\mathrm{A}$ and P$_\mathrm{B}$) are used, where position in the Cartesian axis are considered.  Field  parallel to the $k_{z}$ axis is assumed. } 
\label{Fig:TaAs_WP1}
\end{figure*}

\begin{figure}[t]
\includegraphics[width=0.6\textwidth]{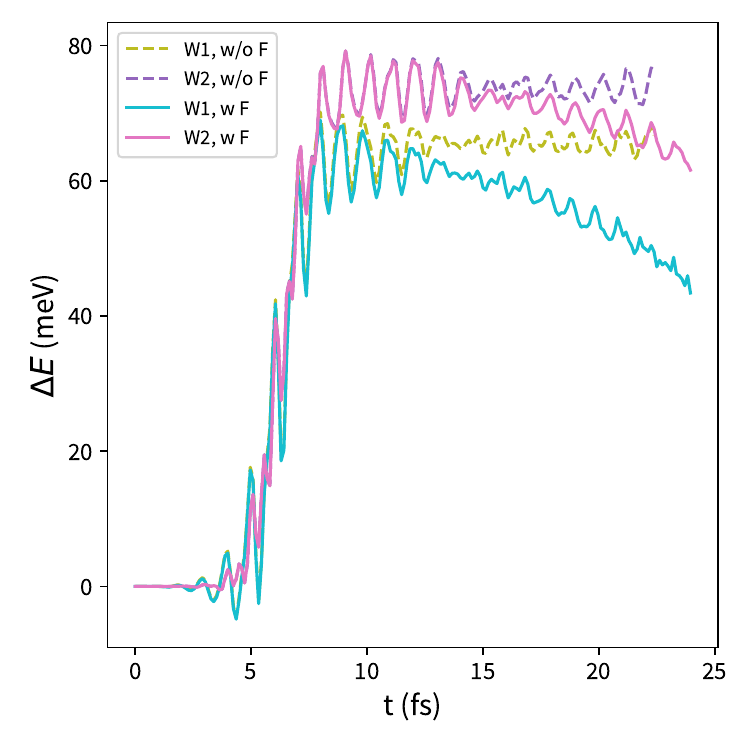}%
\caption{Laser pulse induced change of the WNs energy levels with respect to the application of the Ehrenfest dynamic. (solid line) relaxation included, (dashed line) without relaxation.} 
\label{Fig:WN_energy_evol}
\end{figure}

%
%
Irrespective of the  field strength, qualitatively similar dynamics of the WN positions were detected. The laser pulse induced modifications of the band structure (Fig.~\ref{Fig:band_reco}) introduced shifts and oscillations of WN's positions in the k-space (Fig.~\ref{Fig:TaAs_WP1}) as well as a change of the W1 and W2 energy levels (Fig.~\ref{Fig:WN_energy_evol}). Regarding the k-space position, the largest oscillation occurred during the pulse in the  $k_{z}$-direction, which is parallel to the laser pulse field (Fig.~\ref{Fig:TaAs_WP1}). The immense oscillations can be attributed to the Stark shift. Comparing the induced shift to the applied effective electric field, the WN position displacement tends to follow the direction of the electric field $E_{\rm{z}}$ (Fig.~\ref{Fig:TaAs_WP1}d).
Regardless of the pulse strength (P$_{\rm{A}}$ vs. P$_{\rm{B}}$), the oscillations acquire similar magnitudes due to almost identical vector field amplitudes $A_{\mathrm{z}}$. So, the integral of the $E_{\mathrm{z}}$ field component reached over the half period, driving the shift, is comparable. 
%

Actually the WNs positions do not oscillate  only in the $k_{\mathrm{z}}$ direction, but the oscillations occur in the other directions as well (Fig.~\ref{Fig:TaAs_WP1}).  
For simplicity, Cartesian axes are considered instead of the non-orthogonal reciprocal  axes.
We note, that the oscillations persist after the laser pulse keeping an alike period.
The non-vanishing oscillations result from a system's response to the applied pulse and relaxation of the excited state. It is manifested in the total current (Fig.~\ref{Fig:Jz_current}) showing ongoing charge redistribution as the system tries to reach an equilibrium.

However, a more substantial detected feature is  an induced displacement of the 
{WN mean position} (Figs.~\ref{Fig:WN_overview},~\ref{Fig:TaAs_WP1}). It is propagated during the laser pulse and a residual shift remains even after the pulse.  The displacement is well pronounced in the $k_{\mathrm{x}}$ and $k_{\mathrm{y}}$-directions. Nevertheless, for the $k_{\mathrm{z}}$-direction the onset is overwhelmed by the initial immense oscillations for the laser pulse duration and only residual displacement is noticeable.

We studied the induced displacement for both kinds of WNs.
Regarding the W1 WNs, lying in the $k_{\mathrm{x}}k_{\mathrm{y}}$-plane, a significant residual displacement was observed for both of the in-plane components. Having compared the dynamics of several W1 nodes within the plane (Fig.~\ref{Fig:WN_symmetry}),  the most dominant effect is represented by shifting the WNs positions out of the BZ center ($\Gamma$-point). Meanwhile, the WN pair gets closer as their separation in the k-space decreases. Due to the zero $k_{z}$-component, restricted by symmetries, no residual  shift occurs in this direction.
%
Similarly to  W1 node, the W2 nodes get gathered by the laser pulse (Fig.~\ref{Fig:WN_symmetry}), where the relative change of the nodes' separation correspond to W1 nodes 
(Fig.~\ref{Fig:WP_pair_bands}).

A detailed origin of the WN position shift resulting from the time-dependent band structure reconstruction as depicted in Fig.~\ref{Fig:WN_band_str_reconstr_t}. Since it is hard to handle the shifted Weyl cones, we compared the band structure evolution along the Cartesian axes in the vicinity of the WNs. Regarding the W1 nodes, it is evident that the band structure is almost unchanged along the $k_{\mathrm{z}}$-direction in the WN's vicinity except the energy shift originated from the laser pulse delivered energy. Nearby the WN, the initial band separation and curvature is kept for the selected time step. Lying in the $k_{\mathrm{x}}k_{\mathrm{y}}$-plane, the band structure respects the TaAs symmetry, which does not allow the WN position to shift out of this plane. The WN shift in the  $k_{\mathrm{x}}$- and $k_{\mathrm{y}}$- directions can be explained by a laser-induced separation on the valence and conduction bands appearing towards the $\Gamma$-point. It moves WNs out of the $\Gamma$-point in the $k_{\mathrm{y}}$-coordinate and simultaneously shrink the WNs' separation along the $k_{\mathrm{x}}$ one (Fig.~\ref{Fig:WN_band_str_reconstr_t}). Concerning the W2 nodes, an akin model works. Only in the WN vicinity, the band structure reconstruction along the $k_{\mathrm{y}}$-direction is negligible as the dominant WN displacement resides along the $k_{\mathrm{z}}$-component.

The induced residual WN displacement in the k-space occurs even for much weaker laser fluence.  Although the related electron excitations are not too significant (Fig.~\ref{Fig:Occ_strong_p}b), the induced WNs displacements are quite remarkable (Fig.~\ref{Fig:TaAs_WP1}). They follow previously described behavior and the oscillations observed in the displacement follow the relevant laser pulse frequency. 
Assuming different pulse strengths, the acquired residual displacement seems to be nearly proportional to the square root of the used laser fluence, expressing the amount of energy shinned at the sample. It keeps the residual displacement visible even for the weaker laser pulse with sufficiently reduced fluence.
The relation likely comes from the significant displacements predominantly in two dimensions only.

Regarding the stronger pulse P$_{\rm{A}}$, the maximal magnitude of the residual  displacement $\Delta_{\rm{WN}}$ reached by the W1 nodes is $\Delta_{\rm{WN}} \sim 0.8 \% k_{\rm{c}}$ resp. for the weaker pulse P$_{B}$ $\Delta_{\rm{WN}} \sim 0.25 \% k_{\rm{c}}$. It stands for quite significant modification, which might have a substantial impact on the samples susceptibility (Fig.~\ref{Fig:Susceptibility_Ch31}) .

Along with WN's k-space position modification, the WNs' energy levels are modified by the laser pulse (Fig.~\ref{Fig:WN_energy_evol}). Interestingly  the energy separation of the W1 and W2 WN's is changing as well (Fig.~\ref{Fig:WN_band_str_reconstr_t}), where the laser pulse enhance their separation of about 10 meV. 
%
%
%
%
Importantly, the significant change of the WN's energy separation should be more apparent in the experiment than tiny modification of the WN's positions.

Thanks  to a complex spin texture related to the occurrence of WNs, there might exist signatures in the spin response function (Eq.~\ref{Eq:KResponse})  originating from presence of WNs.
Revealing the change of the WNs separation, the response function  might reflected laser pulse induced WNs dynamics as transitions between W1 and W2 can appear.
For simplicity,  the  nearest W1 and W2 possessing opposite  chirality were chosen (Fig.~\ref{Fig:Susceptibility_Ch31}d). In order to identify the origin of the response features, we considered only a small segment of the  BZ. Assuming the proper q-vector between the W1 and W2 WNs (Fig.~\ref{Fig:WN_kdistance}) and k-points in their vicinity (Fig.~\ref{Fig:Susceptibility_Ch31}d), a  transition (Fig.~\ref{Fig:Susceptibility_Ch31}c) at the energy separation (Fig.~\ref{Fig:WN_edistance}) of the WNs was observed.  Comparison of  response functions for titled q-vector orientation and different k-space segments suggest its relation right to the W1 and W2 transition. We show that the transition follow the WNs' energy separation as proper q-vector is considered. It points that WNs dynamics is reflected also in the spin response.




\subsubsection{Relaxation effects}
So far, we have considered fixed  lattice sites during the time evolution. Therefore, after the laser pulse, the electronic subsystem seems to reach a quasi-equilibrium state and the resulting residual WN displacement (Fig.~\ref{Fig:TaAs_WP1}) as well as  WNs' energy levels (Fig.~\ref{Fig:WN_energy_evol} are almost unchanged with the increasing  time delay.
Thus, allowing for the Ehrenfest dynamics driving the ions out of the ground state equilibrium positions seems to be important. It enables the electron system  to dissipate the acquired energy towards the lattice vibrations and try to restore the initial state. Including the ions motion, the WN dynamics was rather unaffected below $t$=10~fs, when the laser pulse was applied (Figs.~\ref{Fig:TaAs_WP1},~\ref{Fig:WN_energy_evol}). Later, remarkable relaxation processes in the WN's k-space and energy position occur. Their onset is also apparent from laser pulse induced current (Fig.~\ref{Fig:Jz_current}), where the evolution start to deviate from the calculation neglecting the ion motion proving changes in the electron density.

However, relaxation processes are pronounced particularly in WPs energy level changes (Fig.~\ref{Fig:WN_energy_evol}) and their resulting separation (Fig.~\ref{Fig:WN_edistance}). Unlike the calculation without the Ehrenfest dynamics, the WN energy level tends to the original positions after the laser pulse (Fig.~\ref{Fig:WN_energy_evol}). Moreover, a faster relaxation  of W1 WN results in the growth of the WNs energy separation (Fig.~\ref{Fig:WN_edistance}), which is reflected in the change of the resonance position in the response function 
(Fig.~\ref{Fig:Susceptibility_Ch31}).


Besides,  relaxations occur in the WN k-space position, particularly for the W2 nodes and their $k_{x}$ component (Fig.~\ref{Fig:TaAs_WP1}). It exhibits strong reduction of the residual displacement beyond the 15 fs.  Further, an onset of the relaxation is visible in the $k_{y}$ direction for later time. Relaxations of the WN displacement occurs for the W1 nodes as well, e.g. the $k_{y}$, direction. These effect are related to the reconstruction of the band structure (Fig.~\ref{Fig:WN_band_str_reconstr_t}) due to the energy dissipation.


The shown relaxation effects are limited by short calculated time delay. It results from the simple treatment of the time evolution restricting the available time range by the numerical stability and computational demands. The problem might be possibly overcame by using another solver e.g. assuming the self-consistent treatment  in each time  step. It would enlarge the eligible time step length and decrease the computation demands.

Laser induced reduction of the WN k-space separation has been experimentally reported for more complicated WTe$_{2}$ system~\cite{r19_Sie_WTe2_WNshifting} on much longer time scales,  where is explain there by induced shear modes modifying the lattice. Our calculation provide similar effect, describing the laser induced band structure relaxation giving rise shifting of the WN nodes as well as a relaxation due to lattice dynamics.


%% file: nanolett_conclusions.tex
\section{Conclusions} 


To conclude, based on the TD-DFT calculations, this study offers an effective description of the banstructure renormalization  induced by optical pump. Focusing on the TaAs Weyl semimetal, a remarkable dynamics of the presented Weyl nodes was revealed.   Having shined TaAs by a laser pulse,  substantial energy level shift and  displacement of the WNs  was observed.  Interestingly,  the induced changes survive even after the pulse duration. For both types of the existing pairs of WNs in the TaAs system, an induced motion of WNs in the k-space along with long lasting modification of WNs' energy levels and their separation were observed. 


The evidence of the laser pulse  induced WN displacement correspond with the experimental result proving decreasing separation of WN node in another WN material.  We demonstrated the  ions dynamics influence on the  relaxation of the WN dynamics through a energy dissipation to the lattice. 


%% file: nanolett_supportinfo.tex
\subsection{Calculation methods}
%

%

The calculations provided in this  work  were performed within the \emph{Elk} code~\cite{ElkCODE}, the all-electron full-potential  linearised-augmented-plane-wave (LAPW)~\cite{book_SinghNordstrom_LAPW} package. It represents a robust and powerful  open-source tool, which allow us treat the ground state density functional theory (DFT) calculation as well as more advanced feature \emph{i.e.} real-time time-dependent  density functional theory (TDDFT) system evolution~\cite{Elliott2016_TDDFT} and linear response calculations~\cite{Dewhurst2021_magdyn}. It includes our further modification to describe the time-dependent band structure and Weyl node (WN) dynamic.

\subsection{DFT formalism}

Ground state DFT calculations were performed of the 12x12x12 k-mesh, while the exchange correlation potential in the generalized gradient approximation (GGA) of Perdew–Burke-Ernzerhof (PBE)~\cite{r96_Perdew_PBE} type was included. Regarding the TDDFT evolution calculation, the same parameters were use as for the ground state including the adiabatic GGA (AGGA) xc-potential approach.
The experimental crystal structure~\cite{w_Spinger_crystdata,r65_Furuseth_TaAs_cryst} was considered.

%

The ground state in the framework of the Elk code is determined by the common Kohn-Sham (KS) equation~\cite{r18_Baer_Tddft,r65_KohnSham}
\begin{align}
\left( - \nabla^{\rm{2}} + v_{\rm{ext}}(\mathbf{r}) +\int \frac{n(\mathbf{r}^{\prime})}{\vert \mathbf{r} - \mathbf{r}^{\prime} \vert}\mathrm{d}\mathbf{r}^{\prime} +  v_{\rm{xc}}(\mathbf{r})  \right)&\varphi_{\rm{i}}(\mathbf{r})=\varepsilon_{\rm{i}}\varphi_{\rm{i}}(\mathbf{r}) \, ,
\label{Eq:KS_eqs}
\end{align}
where $v_{\rm{ext}}$ is an external potential, $v_{\rm{xc}}$ exchange correlation potential,  $n(\mathbf{r})$ represents the single particle electron density and $\varepsilon_{\rm{i}}$ stands for the eigenenergy of the \textbf{KS} state $\varphi_{\rm{i}}$. We note we are using the Hartree atomic units in expressions.

In the \emph{Elk}, the KS equations (Eq.~\ref{Eq:KS_eqs}) are solved in the following two variation step scheme. 
%
First, only scalar potential and electric field $\mathbf{E}$ are considered:
\begin{align}
        \hat{H}^{\rm{I}} = \hat{T}_{\rm{S}} + \hat{V}_{\rm{ext}} + & \hat{V}_{\rm{C}} + \hat{V}_{\rm{XC}} + \mathbf{E}\cdot \hat{\mathbf{r}} \label{Eq:1stvarstep_H} \\
        \hat{H}^{\rm{I}}\phi_{\rm{i}}^{\rm{I}} &= \epsilon_{\rm{i}}^{\rm{I}}\phi_{\rm{i}}^{\rm{I}} \, ,
        \label{Eq:1stvarstep}
\end{align}
$\hat{T}_{\rm{S}}$ stands for the kinetic term, $\hat{V}_{\rm{ext}}$ is an external potential,  $\hat{V}_{\rm{C}}$ denotes the Coulomb potential and $\hat{V}_{\rm{XC}}$ represents the xc-potential. The last term in the Eq.~\ref{Eq:1stvarstep_H} is the interaction with an external electric field $\mathbf{E}$, where $\hat{\mathbf{r}}$ denotes the position operator. $\phi_{\rm{i}}^{\rm{I}}$ and $\epsilon_{\rm{i}}^{\rm{I}}$  represent the first-variational eigenvectors resp. eigenenergies. 

To cover  relativistic effects and  emergence of Weyl quasipartiles,
considering of the spin-orbit coupling (SOC) is required~\cite{r18_Armitage_WS_DS_rev,r17_Felser_WSrev}. It is included by  means of the scalar relativistic approach together with the external and xc-magnetic fields (${\mathbf{B}}_{\rm{ext}}$ resp. ${\mathbf{B}}_{\rm{xc}}$), and an applied vector potential $\mathbf{A}$   in the second variational step.
%

%
\begin{align}
  &{H}_{\rm{ij}} =        \epsilon_{\rm{i}}^{\rm{I}}\delta_{\rm{ij}} + \\ \nonumber  & +
        \langle \phi_{\rm{i}}^{\rm{I}} \vert \hat{\bm{\sigma}}\cdot\left( \hat{\mathbf{B}}_{\rm{ext}} + \hat{\mathbf{B}}_{\rm{xc}}  \right)  +  \hat{\bm{\sigma}}\cdot\hat{\mathbf{L}} + \mathbf{A}\cdot{\nabla}   \vert   \phi_{\rm{j}}^{\rm{I}}\rangle \, ,
           \nonumber      \\
       & \hat{H}_{\rm{}} \vert \phi_{\rm{i}}^{\rm{II}} \rangle  =   \epsilon_{\rm{i}}^{\rm{II}} \vert \phi_{\rm{i}}^{\rm{II}} \rangle  =    \epsilon_{\rm{i}}^{\rm{II}} \sum_{{\rm{j}}} c_{\rm{j}}^{\rm{II}}  \vert \phi_{\rm{j}}^{\rm{I}} \rangle 
               \label{Eq:H2stvardiag}
\end{align}
Since, generally, a non-collinear magnetism is considered, the second  variational  eigenenergies $\epsilon_{\rm{i}}^{\rm{II}}$ and eigenvectors $\vert \phi^{\rm{II}} \rangle$ are spinors  and  $\hat{\bm{\sigma}}$ stands for Pauli matrices. 
For simplicity, the second variational eigenvectors $\vert \phi^{\rm{II}} \rangle$  diagonalizing total Hamiltonian $H$ (Eq.~\ref{Eq:H2stvardiag}) are represented in the first variational basis  $\vert \phi_{\rm{j}}^{\rm{I}} \rangle$ (Eq.~\ref{Eq:1stvarstep})  by coefficients $c^{\rm{II}}$ (Eq.~\ref{Eq:H2stvardiag}).

%

\subsection{TDDFT formalism \label{Sec:App_TDDFT_formal}}

The time-evolution of the ground state wave functions is considered as a simple direct propagation without self-consistent treatment in time~\cite{r16_Dewhurst_ElkTimeEvol}. An evolution of a KS state $\vert \varphi(t) \rangle $ in time difference $\mathrm{d}t$   reads 
\begin{equation}
   \vert \varphi(t+\mathrm{d}t) \rangle = \hat{U}(t)\vert \varphi(t) \rangle \; ,
   \label{Eq.StateEvol}
\end{equation}
where $\hat{U}(t)$ is the evolutionary operator
\begin{equation}
    \hat{U}(t)= \mathrm{exp}\left[-i \hat{H}(t)\mathrm{d}t\right]
    \label{Eq.UEvol}
\end{equation}
related to the instantaneous Hamiltonian $\hat{H}(t)$ at the time $t$.

Assuming the \textit{velocity gauge}~\cite{r16_Dewhurst_ElkTimeEvol,r16_Forre_VelG,r22_Mattiat_VelG}, we neglect spatial dependencies of the vector potential $\mathbf{A}$ (Eq.~\ref{Eq:VectorPot}) and impose the Coulomb gauge condition $\nabla \cdot \mathbf{A} =0$. Then, in the \textit{dipole approximation} and the second variational basis  $\vert \phi_{\rm{j}}^{\rm{II}} \rangle$ (Eq.~\ref{Eq:H2stvardiag}) , the Hamiltonian matrix elements reads 
\begin{equation}
 \quad   H_{\rm{ij}}(t) = {V}_{\rm{S(ij)}}(t) + T_{\rm{S(ij)}}(0)  -   \mathbf{A}(t) \cdot  \mathbf{P}_{\rm{ij}}(0) \, ,  \label{Eq.TD_Ham}\\
\end{equation}
where ${V}_{\rm{S}}(t)$ denotes Kohn-Sham potential related to   eigenstates   $\vert \phi_{\rm{i}}^{\rm{II}}(t) \rangle$,   $T_{\rm{S(ij)}}$ is the initial kinetic part (Eq.~\ref{Eq:H_Kin_term}) and the final term stands for  the interaction with the external vector potential $\mathbf{A}(t)$ using  the momentum matrix $\mathbf{P}_{ij}$.

\begin{align}
 &T_{\rm{S(ij)}} = \varepsilon_{\rm{i}}^{\rm{II}}\delta_{\rm{ij}}  -  \langle  \phi_{\rm{i}}^{\rm{II}}  \vert {V}_{\rm{S}} \vert \phi_{\rm{j}}^{\rm{II}} \rangle \, , \label{Eq:H_Kin_term}  \\
&\mathbf{P}_{ij}=\int d^3r\,\phi_{i{\bf k}}^{\rm{II}\:*}({\bf r})\left(-i\nabla
    +\frac{1}{4}\left[\vec{\sigma}\times\nabla V_{\rm{S}}({\bf r})\right]\right)
    \phi_{j{\bf k}}^{\rm{II}}({\bf r)} \: , \label{Eq:H_Mom_matr} \\
&  \mathbf{A}(t)= - \int_{0}^{t} \mathbf{E}(\tau)\mathrm{d}\tau \\
&\hbar,c,e=1 \,  \nonumber,
\end{align}

In this work, we considered a time evolution induced by a linearly polarized laser pulse. It is describe by a  vector potential $\mathbf{A}(t)$  constructed from a  sinusoidal wave  modulated with a Gaussian envelope 
function 
\begin{equation}
    \mathbf{A}(t)={\mathbf{A}}_0
    \frac{e^{-(t-t_{\rm{p}})^2/2\sigma^2}}{\sigma\sqrt{2\pi}}
    \sin\left[\omega(t-t_{\rm{p}})+\phi\right]\; 
    \label{Eq:VectorPot}
\end{equation}
parameterized by the vector  amplitude ${\mathbf{A}}_0$, peak time $t_{\rm{p}}$, full-width at half-maximum  $d=2\sqrt{2\ln 2}\sigma$, frequency $\omega$ and phase shift $\phi$.

Diagonalizing the time-dependent Hamiltonian (Eq.~\ref{Eq.TD_Ham}), one obtains third variational vectors $\vert \phi_{\rm{i}}^{\rm{III}} (t) \rangle$, so-called \textit{Houston states}~\cite{r15_Wu_HoustonSt,r86_Krieger_HoustonBasis}, with eigen-energies $\epsilon_{\rm{i}}^{\rm{III}}(t)$
\begin{equation}
    H_{\rm{ii}}(t) \:  \vert \phi_{\rm{i}}^{\rm{III}} (t) \rangle = \epsilon_{\rm{i}}^{\rm{III}}(t) \: \vert \phi_{\rm{i}}^{\rm{III}} (t) \rangle \, .
    \label{Eq.:Houston_states}
\end{equation}
Then,the evolved states are given by a simple formula
\begin{align}
&\vert \phi^{II}_{\rm{i}} (t+\mathrm{d}t) \rangle  \nonumber = \\& = \sum_{i} \mathrm{e}^{ -i \varepsilon_{\rm{j}}^{\rm{III}}(t) \mathrm{d}t} \langle    \phi^{II}_{\rm{i}}(t) \vert \phi^{\rm{III}}_{\rm{j}} (t) \rangle  \:  \vert \phi^{II}_{\rm{i}} (t)\rangle  \: .
\label{Eq.2vec_evol}
    \end{align}





The most straightforward way to track the band structure evolution might be following an evolution of ground state wave functions (Eq.~\ref{Eq:H2stvardiag})  and evaluation of the expectation values of the instantaneous Hamiltonian $\hat{H}(t)$ (Eq.~\ref{Eq.TD_Ham}) related  the time evolved states $\vert \phi_{\rm{i}}^{\rm{II}}(t) \rangle$  as follows
\begin{equation}
    \varepsilon_{\rm{i}}^{\rm{II}}(t)= \langle \phi_{\rm{i}}^{\rm{II}}(t) \vert \hat{H}(t)  \vert \phi_{\rm{i}}^{\rm{II}}(t) \rangle \: .
    \label{Eq.HamTExpV}
\end{equation}
Let's call this approach a ground-stated evolved band structure.

%

Nonetheless, the given approach fails in presence of  inter-band transitions. 
%
In such case, there arises an interchange of contributions to the expansion coefficients 
\begin{equation}
     \vert \phi_{\rm{i}}^{\rm{II}} \rangle (t) = \sum_{j} c_{\rm{j}}^{\rm{III}}(t) \vert \phi_{\rm{j}}^{\rm{III}} \rangle (t)
\end{equation}
 between the coupled bands leading to an energy shift of particular bands. Mixing the expansion coefficients, the related band energies  are being corrupted locally, which  gives rise to  an artificial twisting of the band structure.

Regarding failures of the progressively evolved band structure from ground state, it is more appropriate to determine the band structure based on the time-dependent eigenvalue spectrum (Eq.~\ref{Eq.:Houston_states}) of the instantaneous Hamiltonian $H_{\mathrm{ij}}(t)$ (Eq.\ref{Eq.TD_Ham}) defining system's instantaneous states -- the Houston states $\vert \phi_{\rm{i}}^{\rm{III}} (t) \rangle$.
Their occupancies and character are obviously determined by  projections to the second variational basis $\langle \phi_{\rm{i}}^{\rm{II}}(t) \vert \phi_{\rm{i}}^{\rm{III}}(t) \rangle$, assuming the initial occupancy and character of the initial  states  $\vert \phi_{\rm{i}}^{\rm{II}}(0) \rangle$ (Eq~\ref{Eq.2vec_evol})

\begin{figure}[t]
\includegraphics[width=\textwidth]{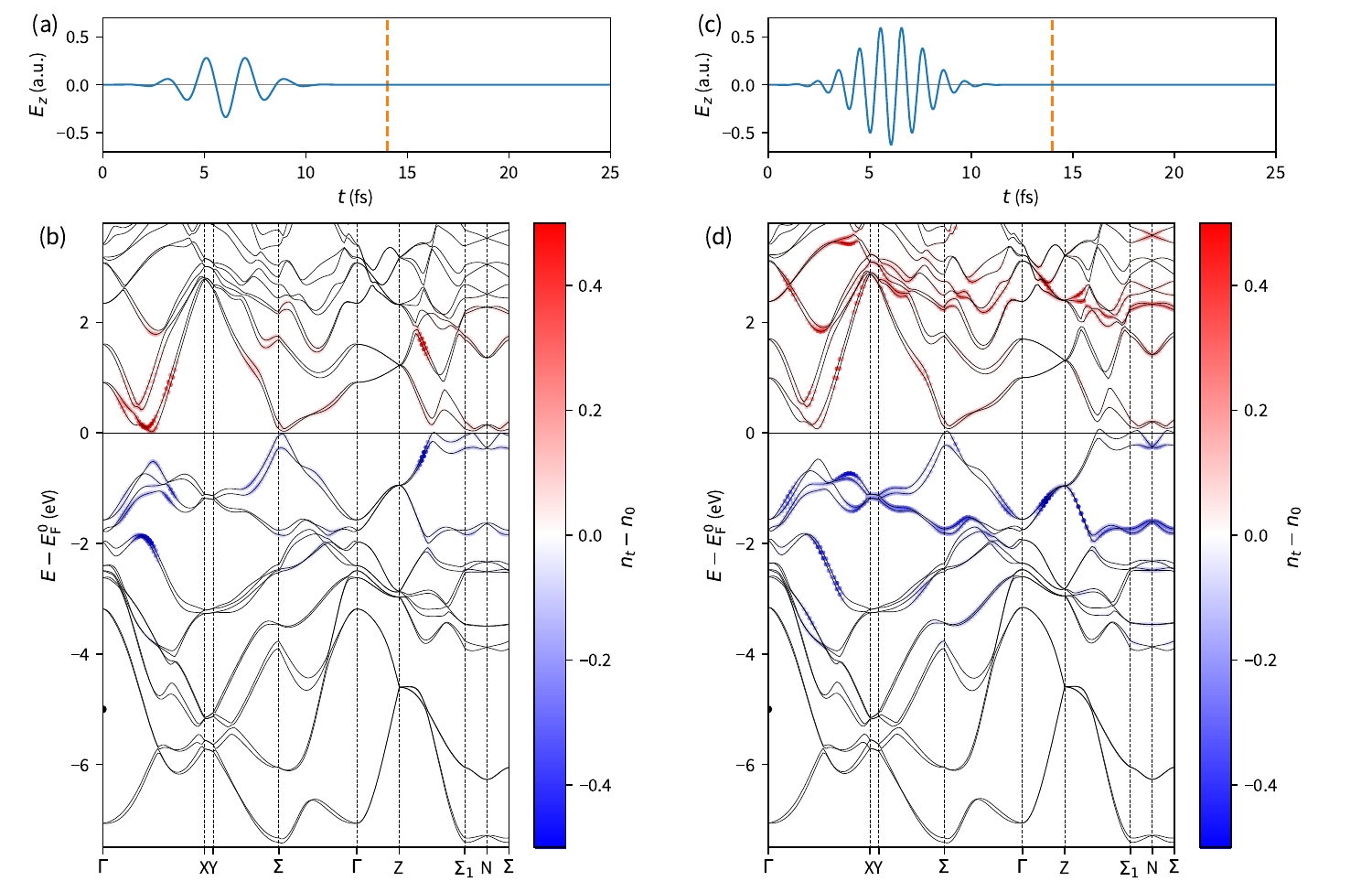}%
\caption{Change of the band occupation after at 14~fs. (a-b) 2.0~eV pulse, (c-d) 3.9~eV pulse. (a,c) Applied electric field. (b,d) Band structure with a depicted change of the occupancy with the respect to the ground state.  The energy is scaled to the ground state Fermi level $E_{\mathrm{F}}^{0}$} 
\label{Fig:Occ_strong_p}
\end{figure}

To study an evolution of the occupation numbers or band character along a selected k-path, an auxiliary k-set representing an arbitrary k-path has to be included as the evolution at particular k-points has to be tracked  from the initial step (Eq.~\ref{Eq.StateEvol}) (Fig.~\ref{Fig:Occ_strong_p}).
Possibly, an auxiliary k-mesh can be avoided if only  a band structure spectrum is desired. Storing the instantaneous charge density $n(\mathbf{r},t)$, magnetic spin density $m(\mathbf{r},t)$  and  Kohn-Shame potential  ${V}_{\rm{S}}(t)$, the related instantaneous Hamiltonian $\hat{H}(t)$ (Eq.~\ref{Eq.TD_Ham})  for the applied vector field $\mathbf{A}(t)$.
can be restored and used to determine eigenvalues $\varepsilon_{\rm{i}}^{\rm{III}}(t)$ along an arbitrary k-path.

It is worthy  mention that the   densities $n(\mathbf{r},t)$ and Kohn-Shame potential  ${V}_{\rm{S}}(t)$ were obtained for the transient occupations at the time $t$.
Therefore, for the applied vector field $\mathbf{A}(t)$,  they directly determine the instantaneous states  (Eq.~\ref{Eq.:Houston_states}) at the time $t$.
All the potential and densities are related to an appropriate occupation arising from the TDDFT calculations
We show that the restored eigen values corresponds to the directly evolved on the auxiliary  k-mesh (Fig.~\ref{Fig:Occ_strong_p}). 


In our calculation, we applied a linearly polarized pulse along the $k_{z}$ axis. It is the simplest choice as the field is parallel to the high-symmetry crystallographic z-axis and does not break the perpendicular plane symmetry. Therefore,  related time evolution calculations are the most feasible ones from the point of view of the computational demands.

\begin{figure}[t]
\includegraphics[width=0.6\textwidth]{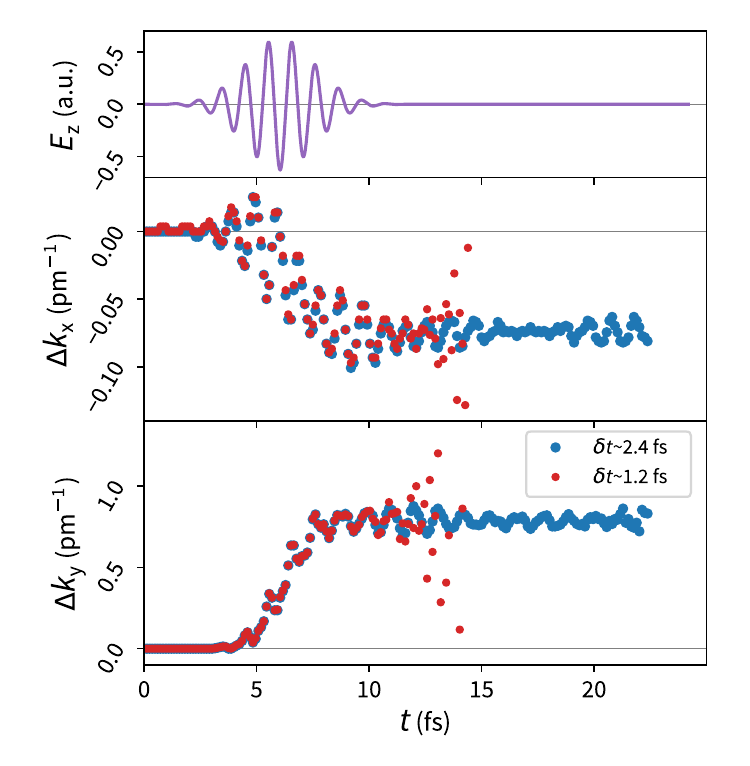}%
\caption{Extracted WN1 position for different TDDFT time step length $\delta t$. Laser pulse P$_{\mathrm{A}}$ is considered.} 
\label{Fig:WN1_dt_fail}
\end{figure}

Originally, we had performed the TD-DFT calculations using a time step $\delta t$=0.10~a.u. ($\sim$ 2.4~as) and laser pulse  width  FWHM$\sim${7.3~fs}. Nevertheless, even for the not large studied TaAs system, the TD-FT evolution suffered from numerical instability after a few fs. It is manifested  by sudden scattering in the WN  position evolution in the Fig.~\ref{Fig:WN1_dt_fail}. This feature was sensitive to the time step length and it could be removed by a shorter time step. Therefore, we squeezed the laser pulse width (FWHM$\sim$3.6~fs) and considered a shorter time $\delta t=$0.05 a.u.), which provided us a longer time window to observed WP dynamics after the laser pulse.

\begin{figure}[t]
\includegraphics[width=0.8\textwidth]{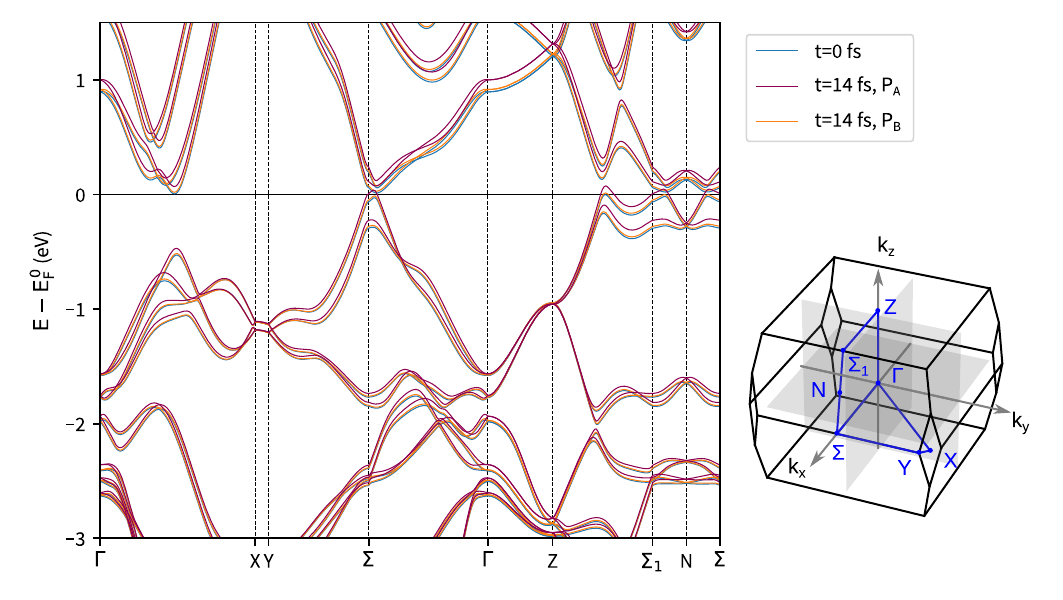}%
\caption{Band structure  renormalization. Comparison of the effect of the strong P$_{\rm{A}}$  and weak P$_{\rm{B}}$ laser pulses. Band are depicted with the respect to the ground state Fermi level $E_{\mathrm{F}}^{\rm{0}}$.} 
\label{Fig:band_reco}
\end{figure}

Two different strength of the linearly polarized laser pulses were used to scale the observed effects. The stronger one with the laser fluence of 10~mJ/cm$^{\rm{2}}$ with the energy 3.9 eV (peak 7.4 $\cdot 10^{11}$ W/cm$^2$) -- called P$_{\rm{A}}$ and weaker possessing fluence 0.3~mJ/cm$^{\rm{2}}$ and 2.0 eV  (peak 2.2 $\cdot 10^{11}$ W/cm$^2$)  -- called P$_{\rm{B}}$ (Fig.~\ref{Fig:band_reco}) 

\begin{figure}[t]
\includegraphics[width=0.7\textwidth]{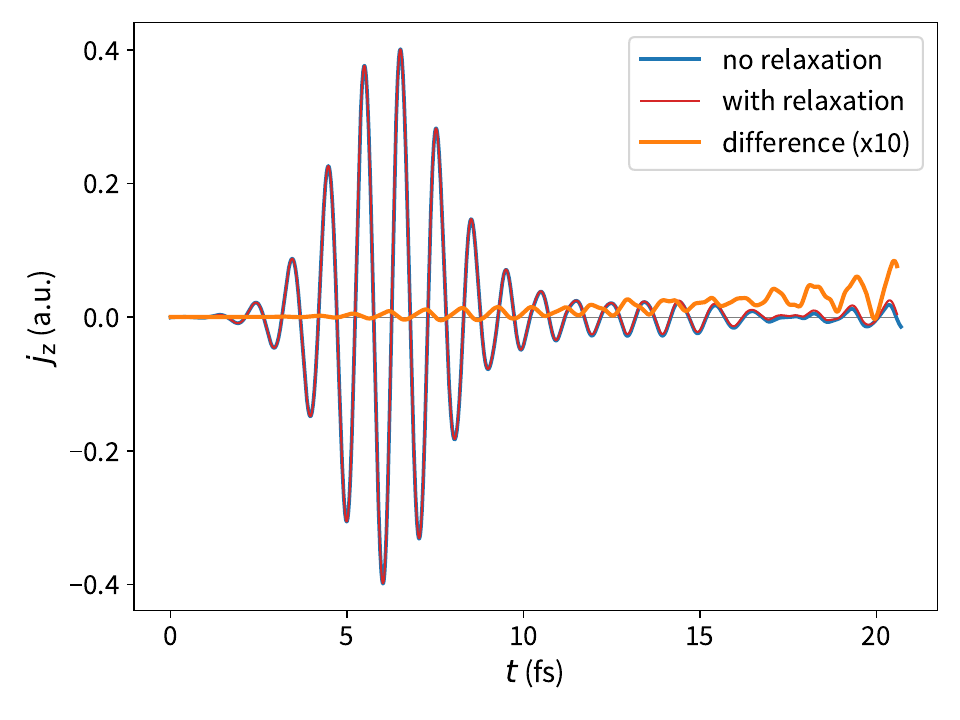}
\caption{Time-dependent laser pulse induced total current. The $j_{z}$ component parallel to the field is depicted.  (red) current with atomic sites relaxation, (blue) current without  relaxation , (orange) difference magnified by the factor 10.  }
\label{Fig:Jz_current}
\end{figure}

Initially, the mentioned experimental crystal structure~\cite{w_Spinger_crystdata,r65_Furuseth_TaAs_cryst} was considered neglecting the laser induced ions motion through the time evolution. Later, to examine the effect of the lattice relaxation, Ehrenfest dynamic was included in the TDDFT calculation.  A simple approach was applied to cover modifications of the nuclear Coulomb potential caused by atomic displacement. It includes extra contribution to the Coulomb potential based on the gradient of the initial nuclear potential and displacement arising from previously evaluated time-dependent inter-atomic forces~\cite{r91_Yu_LAPW_forces,r21_Sharma_tddft_FePt} (Fig.~\ref{Fig:Jz_current}).

\FloatBarrier
\newpage

\subsection{Weyl nodes' dynamics}

Weyl nodes (WN) represent  monopoles and antimonopoles of the Berry curvature~\cite{r18_Armitage_WS_DS_rev,r17_Felser_WSrev,r17_Wang_WSM_tranport_rev,r21_Xie_TopSemM_rev}. Their presence is characterized by a non-vanishing topological invariant so-called Chern number $C$~\cite{r18_Armitage_WS_DS_rev}. At Weyl nodes, it acquires non-zero values depending on the vortex character. The Chern number is defined by means of the total Berry flux $\mathcal{F}(\mathbf{k})$  over a close surface in the k-space which results in a gauge invariant variable~\cite{r05_Takahiro_DiscrChN,r18_Armitage_WS_DS_rev,r17_IFF48_LNotes,r06_Sheng_ChN}
\begin{equation}
    C=\frac{1}{2\pi}\oint  \mathrm{d}\mathbf{k} \: \mathcal{F}(\mathbf{k}).
\end{equation}
%

Assuming a 2D k-space, the Chern number $C$ can be also attributed to the phase $\gamma$~\cite{r05_Takahiro_DiscrChN,r20_Wang_ChN,r20_Zhao_ChN} pick up during the \textit{parallel transport}~\cite{r10_Di_Bphase_rev,r22_Bradlyn_PhysLectNotes}  along a close loop as follows
\begin{equation}
    C=\frac{1}{2\pi} \gamma .
    \label{Eq.Chern_gamma}
\end{equation}

%
Considering  fine discrete  k-mesh and  single band, the total phase difference $\gamma$ along a close loop represented by a set of k-points \{$\mathbf{k}_{\rm{1}}$, $\mathbf{k}_{\rm{2}}$, ..., $\mathbf{k}_{\rm{M}}$, $\mathbf{k}_{\rm{1}}$\}  reads~\cite{r05_Takahiro_DiscrChN,r20_Wang_ChN,r20_Zhao_ChN} 
\begin{align}
    \gamma = \mathrm{Im} \log \left[ \prod_{i=1}^{M} U_{i,i+1} \right] ; M+1 \equiv 1 \: , \label{Eq:gamma_Usegments} \\
    U_{i,i+1} = \frac{\langle \varphi(\mathbf{k}_{i}) \vert \varphi(\mathbf{k}_{i+1}) \rangle}{\vert  \langle \varphi(\mathbf{k}_{i}) \vert \varphi(\mathbf{k}_{i+1}) \rangle \vert} \: ,
\end{align}
where $U$ is a link variable defined by an overlap of the wave function $\varphi$ at the ends of the segment $i$. Importantly, unlike the phase difference along particular segments, the total phase difference $\gamma$ is gauge invariant quantity and represents an observable.

Regarding a larger system with N occupied band, the link variable become a NxN matrix $\mathbf{U}$ with   elements defined as follows~\cite{r05_Takahiro_DiscrChN,r20_Zhao_ChN,r19_Ivanov_WPminimg}
\begin{equation}
    U^{i,i+1}_{mn} =\frac{\langle \varphi(m,\mathbf{k}_{i}) \vert \varphi(n,\mathbf{k}_{i+1}) \rangle}{\vert  \langle \varphi(m,\mathbf{k}_{i}) \vert \varphi(n,\mathbf{k}_{i+1}) \rangle \vert} \:
\end{equation}
where $m,n$ denote band indices. Then, the total phase difference $\gamma$ reads
\begin{equation}
    \gamma = \mathrm{Im} \log \Bigg\{ \mathrm{det} \left[ \prod_{i=1}^{M} U_{i,i+1} \right] \Bigg\}  ; M+1 \equiv 1 \, .
\end{equation}

Considering bulk system, the mentioned approach can be apply  to determine  the Berry flux flowing through a closed loop. Then, Weyl nodes'  presence and their positions can be traced by searching for discontinuities in the Berry flux $\mathcal{F}(\mathbf{k})$ along  different directions in the k-space~\cite{r19_Ivanov_WPminimg,r17_Gresch_Z2pack}.


In general, in this paper, we integrated along squared loop with an edge size down to $l \sim 4 \cdot 10^{-2} \mathrm{\AA}^{-1}$ and k-resolution down to $\Delta k \sim10^{-4} \mathrm{\AA}^{-1}$.
We considered  the phase integration  over the 84 lowest-lying bands representing the occupied states in the ground state.

We note that despite the laser-induced excitations over the ground state Fermi level $E_{\mathrm{F}}^{\mathrm{0}}$ (Fig.~\ref{Fig:Occ_strong_p}), the same number of bands were used in the evaluation of the WNs at $t>0$ as those bands are still predominantly occupied.

\begin{figure}[t]
\includegraphics[width=0.6\textwidth]{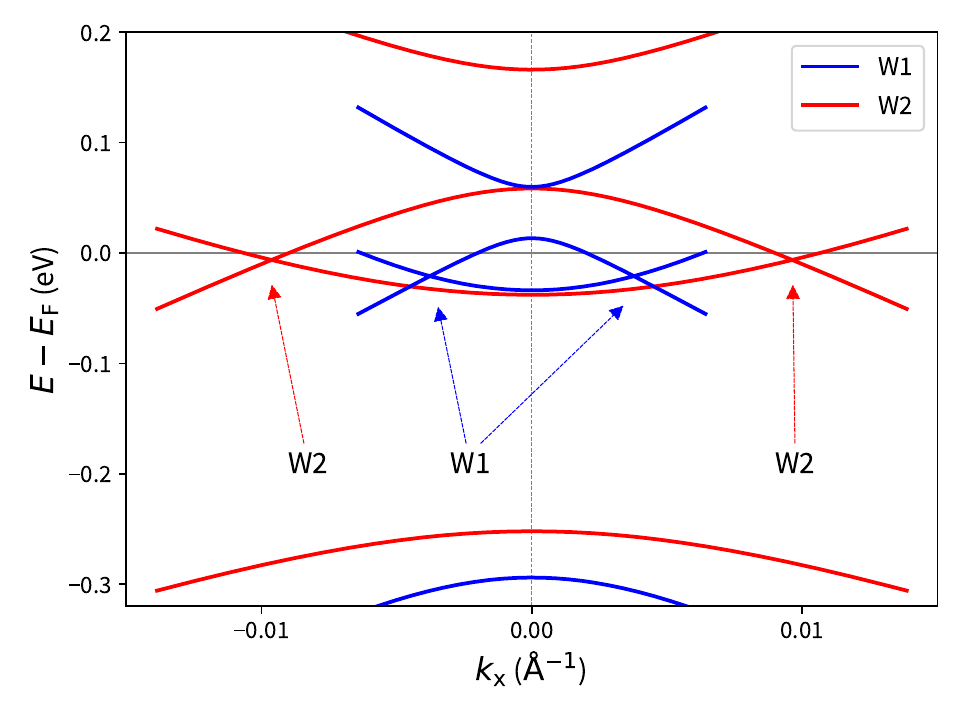}%
\caption{ Ground state  Weyl nodes in TaAs. Band structure  in the vicinity of (blue) W1 and (red) W2 Weyl nodes along line connecting Weyl nodes pairs  are depicted.  } 
\label{Fig:WP_pair_bands}
\end{figure}

\begin{figure}[t]
\includegraphics[width=0.49\textwidth]{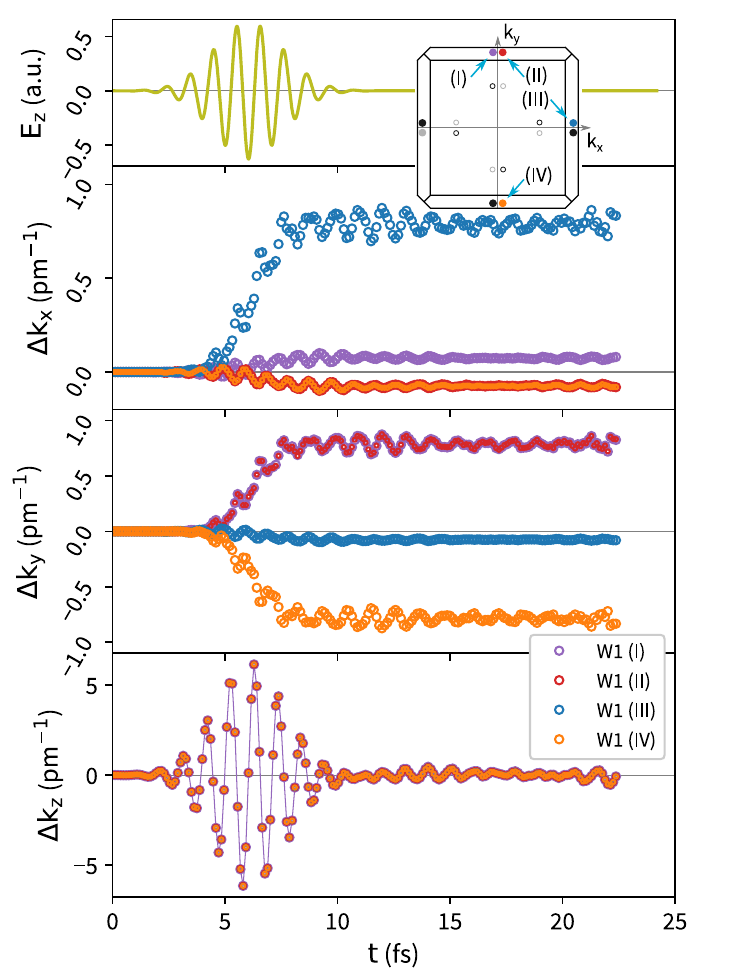}%
\hspace{1pt}
\includegraphics[width=0.49\textwidth]{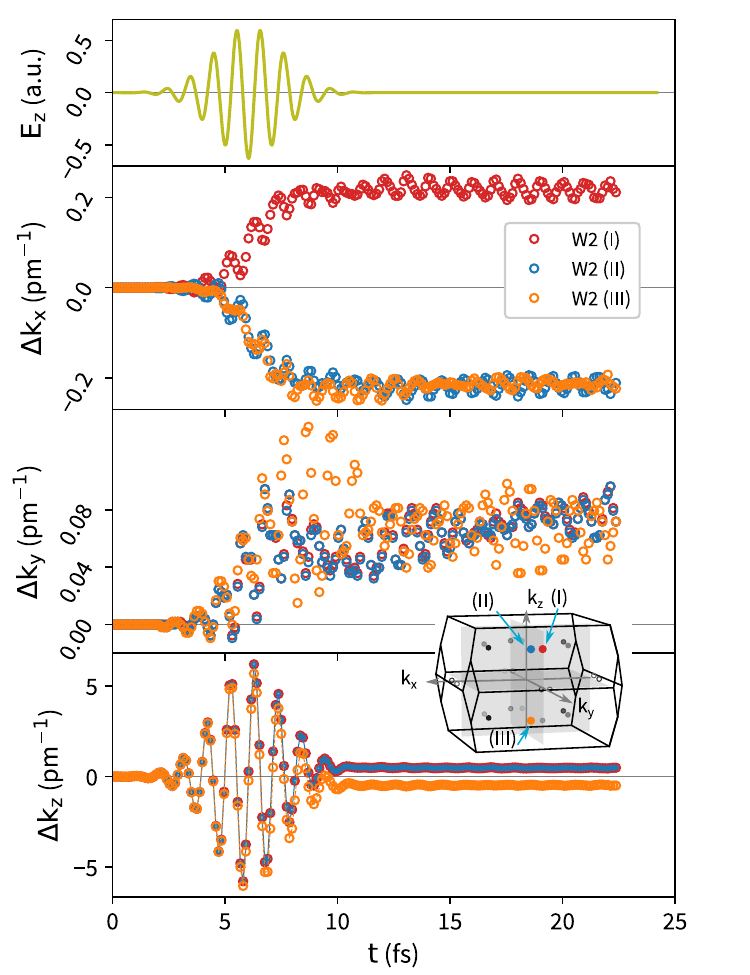}%

\caption{Comparison of time-dependent dynamics at different Weyl nodes. (left) W1 nodes, (right) W2 nodes. Laser pulse P$_{\mathrm{A}}$ is considered.} 
\label{Fig:WN_symmetry}
\end{figure}


\begin{figure}[t]
\includegraphics[width=0.6\textwidth]{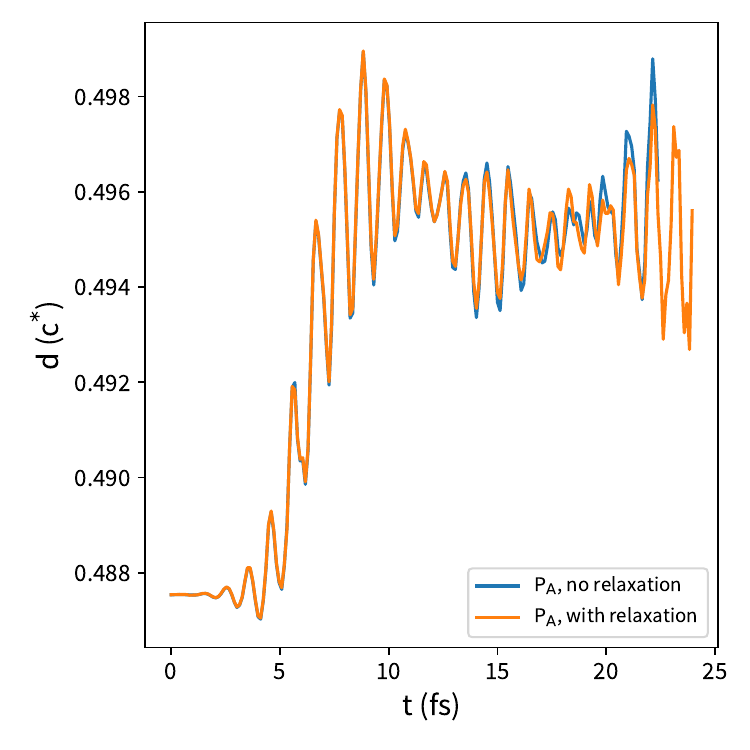}%
\caption{Laser pulse induced change of the k-space distance between the nearest W1 and W2 Weyl nodes of the different chirality. Both data for the evolution with and without the atomic site relaxation are depicted. Laser pulse P$_{\mathrm{A}}$ is considered.} 
\label{Fig:WN_kdistance}
\end{figure}

\begin{figure}[t]
\includegraphics[width=0.6\textwidth]{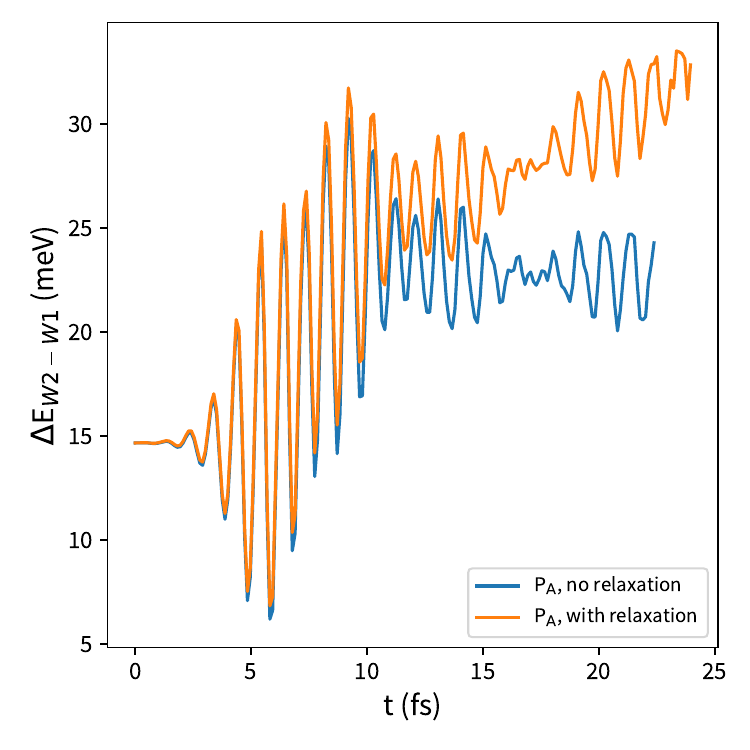}%
\caption{Laser pulse induced change of the energy distance between the  W1 and W2 Weyl nodes with respect to the applied relaxation. Laser pulse P$_{\mathrm{A}}$ is considered.} 
\label{Fig:WN_edistance}
\end{figure}

\begin{figure}[t]
\includegraphics[width=0.6\textwidth]{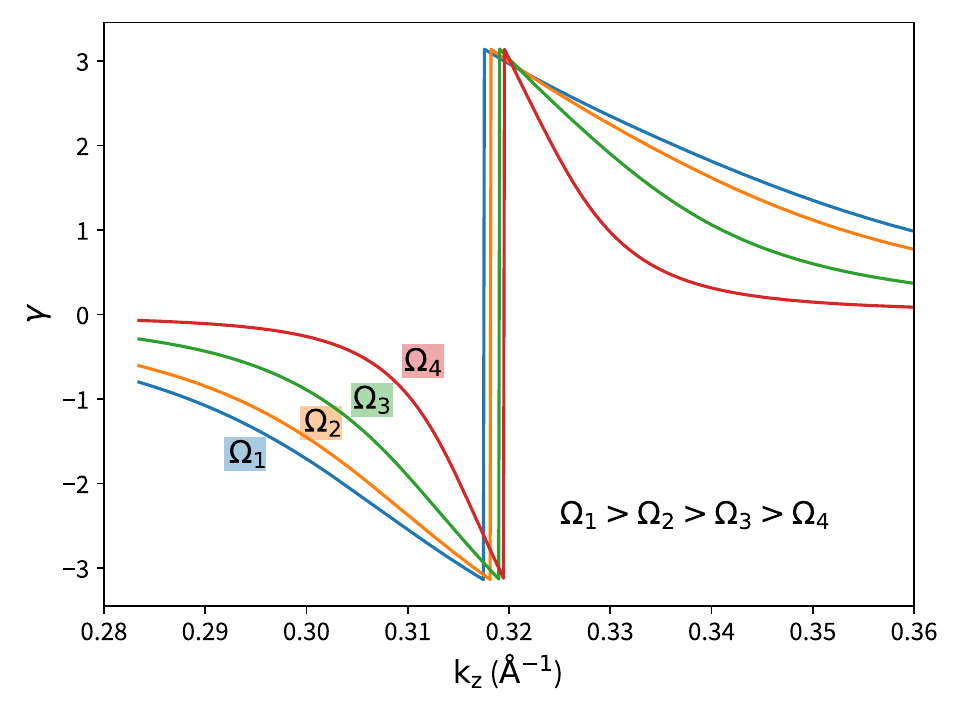}%
\caption{The effect of the Wilson loop size on the integrated phase as a function of the $k_{z}$ position. $\Omega_{i}$ denotes the area of the loop.} 
\label{Fig:WN_loopsize_effects}
\end{figure}

\begin{figure*}[b]
\includegraphics[width=1.\textwidth]{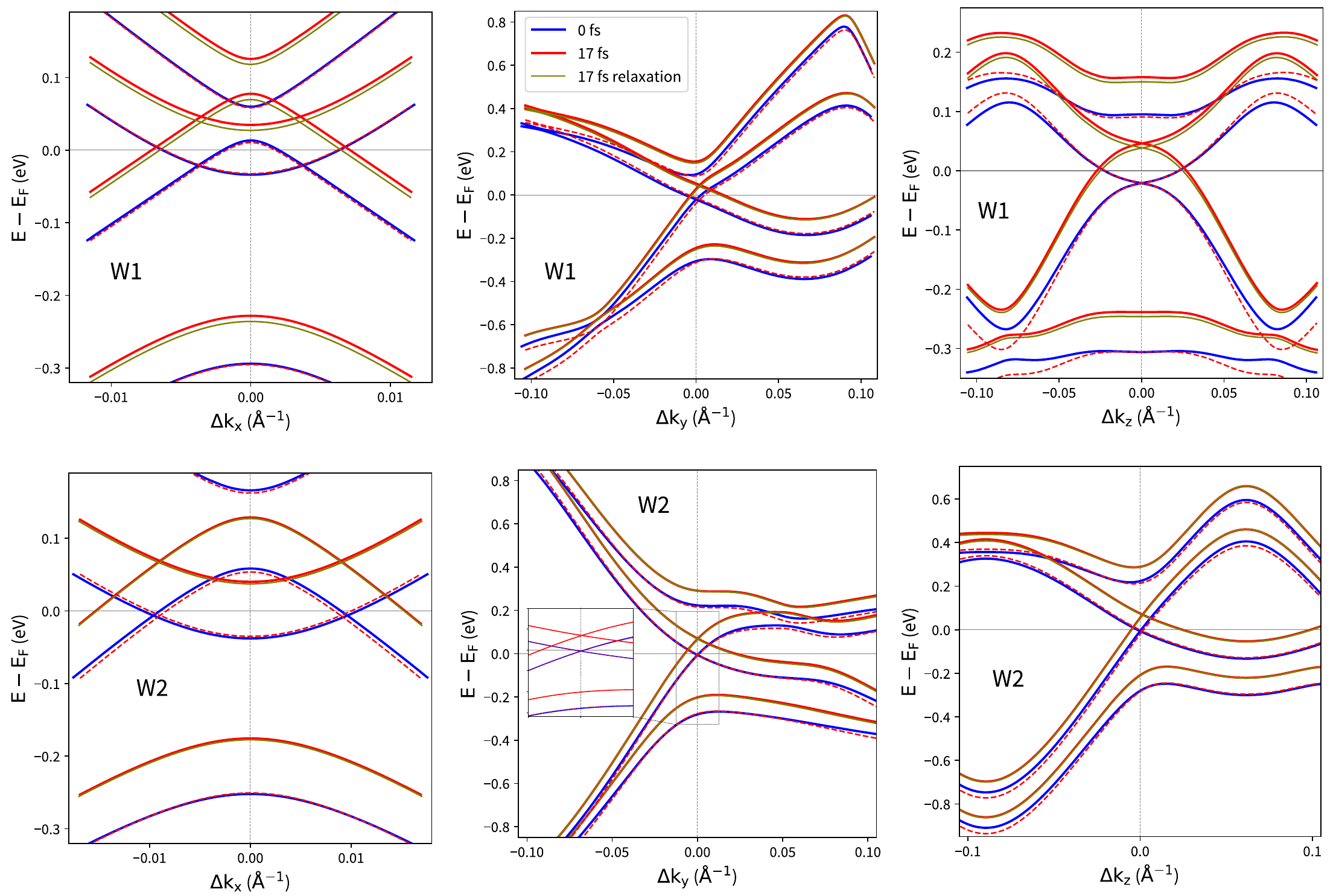}%
\caption{Band reconstruction induced WN position shift. Band structure in the vicinity of Weyl nodes at the (solid blue) t=0~fs and (solid red) t=17~fs are compared, where k-axes are centered at the initial WN position. For clarity, (dashed line) time evolved band structures shifted by the energy difference at the WN with the respect to the initial state are depicted. The P$_{A}$-pulse along the $k_{z}$-direction is considered. The positions of I and II WNs  as depicted in the Fig.~\ref{Fig:WN_symmetry}  are employed.    } 
\label{Fig:WN_band_str_reconstr_t}
\end{figure*}

\FloatBarrier

\subsection{Spin dependent response function}

\begin{figure*}[t]
\includegraphics[width=\textwidth]{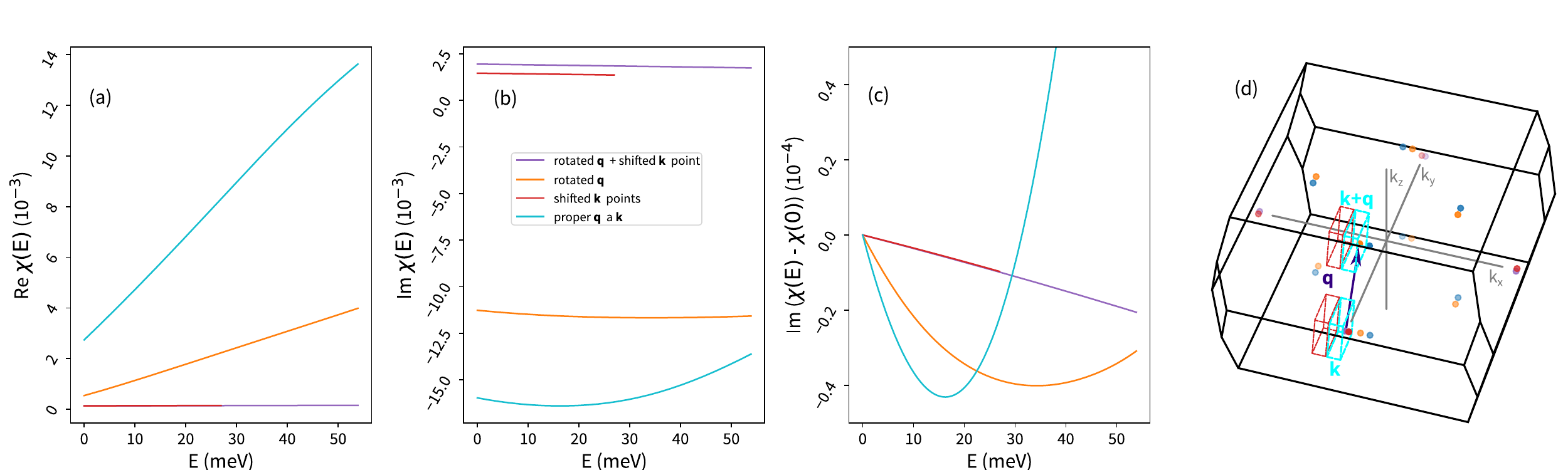}%
\caption{Spin dependent response function. (a) Real part of the response function. (b) Imaginary part of the response function. (c) Change of the imaginary part. (cyan lines) Response related to the  q-vector $\mathbf{q}$ pointing between W1 and W2 WNs of  different chirality and k-points in the vicinity of the WNs. (red lines) Response for  k-points out of the WNs. (orange lines) Response calculated for tilted q-vector $\mathbf{q}$ . (purple lines) Combination of tilted q-vector $\mathbf{q}$ and  k-points shifted out of the WNs (d) Sketch of the considered  q-vector and  k-point range. Cyan boxes restrict assumed  k-points around WNs, whereas  red boxes denotes k-point out of the WNs(Compare Eq.~\ref{Eq:KResponse})}  
\label{Fig:Susceptibility_Ch31}
\end{figure*}

The occurrence of the WNs is related to the presence of chiral state and complex spin structure. 
The studied TaAs Weyl semimetal posses two set  of the WNs, W1 resp W2, lying at distinct energy levels. The energy separation between W2 and W1 WNs is roughly 15 meV in the ground state (Fig.~\ref{Fig:WP_pair_bands}) and it is enlarged by the applied laser pulse (Fig.~\ref{Fig:WN_edistance}). Then, a inter-band transitions between those WNs at distinct energies before and after laser pulse might be observed. To identify them, we evaluated the spin dependent KS response function defined in the real space and frequency domain as follows~\cite{r96_Petersilka_response,r20_Dewhurst_MagDischro}

\begin{align}
    &\chi_{\alpha\beta,\alpha^{\prime}\beta^{\prime}} (\mathbf{r},\mathbf{r}^{\prime},\omega)  \equiv \frac{\partial \rho_{\alpha\beta } (\mathbf{r},\omega) }{\partial \nu_{\alpha^{\prime}\beta^{\prime}} (\mathbf{r}^{\prime},\omega)} = \\ 
    & = \frac{1}{N_{k}}  \sum_{i\mathbf{k},j\mathbf{k}^{\prime}}  \left( f_{i\mathbf{k}} - f_{j\mathbf{k}^{\prime}} \right)  \frac{\langle i\mathbf{k} \vert \hat{\rho}_{\alpha\beta } (\mathbf{r}) \vert j\mathbf{k}^{\prime}\rangle  \langle  j\mathbf{k}^{\prime}   \vert \hat{\rho}_{\alpha^{\prime}\beta^{\prime} } (\mathbf{r}^{\prime}) \vert i\mathbf{k} \rangle}{\omega  + (\varepsilon_{i\mathbf{k}} - \varepsilon_{j\mathbf{k}^{\prime}} ) + i\eta }, \nonumber
\end{align}
where $\alpha$, $\beta$ stands for spin coordinates, $\rho$ is the spin-density, $\nu$ denotes the Kohn-Sham potential 
$N_{k}$ is number of k-points and  $f_{i\mathbf{k}}$ is the occupancy of the state $i$ at the k-point  $k$. $\vert i\mathbf{k} \rangle$ denotes KS states with related eigen-energies   $\varepsilon_{i\mathbf{k}}$, and $\eta$ is a small real positive number 

Applying the Fourier transformation, the response function in the reciprocal space reads
\begin{align}
    &\chi_{\alpha\beta,\alpha^{\prime}\beta^{\prime}} (\mathbf{G},\mathbf{G}^{\prime},\mathbf{q},\omega) = \label{Eq:KResponse}  \\
    & = \frac{1}{\Omega  N_{k}} \sum_{i\mathbf{k},j\mathbf{k}+\mathbf{q}}  \left( f_{i\mathbf{k}} - f_{j\mathbf{k}+\mathbf{q}} \right) 
    \frac{\left[Z_{i\mathbf{k},j\mathbf{k}+\mathbf{q}}^{\alpha\beta}(\mathbf{G})\right]^{*} \:   Z_{i\mathbf{k},j\mathbf{k}+\mathbf{q}}^{\alpha^{\prime}\beta^{\prime}}(\mathbf{G}^{\prime})}{\omega  + (\varepsilon_{i\mathbf{k}} - \varepsilon_{j\mathbf{k}+\mathbf{q}} ) + i\eta } ; \nonumber \\
    &Z_{i\mathbf{k},j\mathbf{k}+\mathbf{q}}^{\alpha\beta}(\mathbf{G}) = \int \mathrm{d}^{3}r \, \mathrm{e}^{i(\mathbf{G}+\mathbf{q})\cdot\mathbf{r}} \varphi^{*}_{j\mathbf{k}+\mathbf{q},\alpha}(\mathbf{r})\varphi_{i\mathbf{k},\beta}(\mathbf{r}) , 
\end{align}
where $\mathbf{G}$ is the reciprocal lattice vector and $\Omega$ is the reciprocal volume.

 To determine whether the response feature originates from the WNs itself, we considered in our calculation only a small segment of the k-space containing the WPs. In the (Fig.~\ref{Fig:Susceptibility_Ch31}), the real and imaginary part of the $\chi_{zx}$ response component denoting the response of the $m_{z}$ component of  magnetization to the change of the $B_{x}$ component of the  magnetic field are depicted. We calculated the response for different part of the k-space and different q-vector orientation.
 Choosing the q-vector pointing in between the WNs and k-points in their vicinity (Fig.~\ref{Fig:Susceptibility_Ch31}d), a resonance in the imaginary part at the energy separation of the WNs was observed indicating a transition between those WNs (Fig.~\ref{Fig:Susceptibility_Ch31}). The resonance's position correspond to the WNs energy separation not only in the ground state calculation, but follow the band structure reconstruction and modification of the WNs energy levels (Figs.~\ref{Fig:WN_edistance}.~\ref{Fig:WN_energy_evol}). 
 Varying of the q-vector or the employed segment of the k-space cancelled the observed transitions.
 Tilting the q-vector in the $k_{y}k_{z}$-plane led to a smearing of the observed resonance. Moreover, assuming k-points out of the WNs in the response function calculation brought a change of the sign of the imaginary part and possessing nearly a linear character indicating no transitions. It suggest close relation of the observed transitions and presence of WNs with given energy separation and k-space position.

 Besides the energy level shifts of the WNs (Fig.\ref{Fig:WN_energy_evol}), it offers a possibility to  detected the laser pulse induced modification of  the WNs in the time (Fig.~{\ref{Fig:TaAs_WP1}) as the their relative position in the k-space (Fig.~\ref{Fig:WN_kdistance}) as well as energy separation (Fig.~\ref{Fig:WN_edistance}) change.

 }

%% file: ms.bbl
\providecommand{\noopsort}[1]{}\providecommand{\singleletter}[1]{#1}%
\providecommand{\latin}[1]{#1}
\makeatletter
\providecommand{\doi}
  {\begingroup\let\do\@makeother\dospecials
  \catcode`\{=1 \catcode`\}=2 \doi@aux}
\providecommand{\doi@aux}[1]{\endgroup\texttt{#1}}
\makeatother
\providecommand*\mcitethebibliography{\thebibliography}
\csname @ifundefined\endcsname{endmcitethebibliography}  {\let\endmcitethebibliography\endthebibliography}{}
\begin{mcitethebibliography}{51}
\providecommand*\natexlab[1]{#1}
\providecommand*\mciteSetBstSublistMode[1]{}
\providecommand*\mciteSetBstMaxWidthForm[2]{}
\providecommand*\mciteBstWouldAddEndPuncttrue
  {\def\EndOfBibitem{\unskip.}}
\providecommand*\mciteBstWouldAddEndPunctfalse
  {\let\EndOfBibitem\relax}
\providecommand*\mciteSetBstMidEndSepPunct[3]{}
\providecommand*\mciteSetBstSublistLabelBeginEnd[3]{}
\providecommand*\EndOfBibitem{}
\mciteSetBstSublistMode{f}
\mciteSetBstMaxWidthForm{subitem}{(\alph{mcitesubitemcount})}
\mciteSetBstSublistLabelBeginEnd
  {\mcitemaxwidthsubitemform\space}
  {\relax}
  {\relax}

\bibitem[Armitage \latin{et~al.}(2018)Armitage, Mele, and Vishwanath]{r18_Armitage_WS_DS_rev}
Armitage,~N.~P.; Mele,~E.~J.; Vishwanath,~A. Weyl and Dirac semimetals in three-dimensional solids. \emph{Rev. Mod. Phys.} \textbf{2018}, \emph{90}, 015001\relax
\mciteBstWouldAddEndPuncttrue
\mciteSetBstMidEndSepPunct{\mcitedefaultmidpunct}
{\mcitedefaultendpunct}{\mcitedefaultseppunct}\relax
\EndOfBibitem
\bibitem[Yan and Felser(2017)Yan, and Felser]{r17_Felser_WSrev}
Yan,~B.; Felser,~C. Topological Materials: Weyl Semimetals. \emph{Annual Review of Condensed Matter Physics} \textbf{2017}, \emph{8}, 337--354\relax
\mciteBstWouldAddEndPuncttrue
\mciteSetBstMidEndSepPunct{\mcitedefaultmidpunct}
{\mcitedefaultendpunct}{\mcitedefaultseppunct}\relax
\EndOfBibitem
\bibitem[Wang \latin{et~al.}(2017)Wang, Lin, Wang, Yu, and Liao]{r17_Wang_WSM_tranport_rev}
Wang,~S.; Lin,~B.-C.; Wang,~A.-Q.; Yu,~D.-P.; Liao,~Z.-M. Quantum transport in Dirac and Weyl semimetals: a review. \emph{Advances in Physics: X} \textbf{2017}, \emph{2}, 518--544\relax
\mciteBstWouldAddEndPuncttrue
\mciteSetBstMidEndSepPunct{\mcitedefaultmidpunct}
{\mcitedefaultendpunct}{\mcitedefaultseppunct}\relax
\EndOfBibitem
\bibitem[Sie \latin{et~al.}(2019)Sie, Nyby, Pemmaraju, Park, Shen, Yang, Hoffmann, Ofori-Okai, Li, Reid, Weathersby, Mannebach, Finney, Rhodes, Chenet, Antony, Balicas, Hone, Devereaux, Heinz, Wang, and Lindenberg]{Sie:2019cf}
Sie,~E.~J. \latin{et~al.}  {An ultrafast symmetry switch in a Weyl semimetal}. \emph{Nature} \textbf{2019}, \emph{565}, 61--66\relax
\mciteBstWouldAddEndPuncttrue
\mciteSetBstMidEndSepPunct{\mcitedefaultmidpunct}
{\mcitedefaultendpunct}{\mcitedefaultseppunct}\relax
\EndOfBibitem
\bibitem[Schäpers(2017)]{r17_IFF48_LNotes}
Schäpers,~T. In \emph{{T}opological {M}atter - {T}opological {I}nsulators, {S}kyrmions and {M}ajoranas}; Blügel,~S., Mokrousov,~Y., Ando,~Y., Eds.; Schriften des Forschungszentrums Jülich. Reihe Schlüsseltechnologien / Key Technologies; Forschungszentrum Jülich GmbH Zentralbibliothek, Verlag: Jülich, 2017; Vol. 139; p getr. Zählung\relax
\mciteBstWouldAddEndPuncttrue
\mciteSetBstMidEndSepPunct{\mcitedefaultmidpunct}
{\mcitedefaultendpunct}{\mcitedefaultseppunct}\relax
\EndOfBibitem
\bibitem[Burkov and Balents(2011)Burkov, and Balents]{r11_Burkov_WSM_prediction}
Burkov,~A.~A.; Balents,~L. Weyl Semimetal in a Topological Insulator Multilayer. \emph{Phys. Rev. Lett.} \textbf{2011}, \emph{107}, 127205\relax
\mciteBstWouldAddEndPuncttrue
\mciteSetBstMidEndSepPunct{\mcitedefaultmidpunct}
{\mcitedefaultendpunct}{\mcitedefaultseppunct}\relax
\EndOfBibitem
\bibitem[Murakami(2007)]{r07_Murakami_3Dgapless}
Murakami,~S. Phase transition between the quantum spin Hall and insulator phases in 3D: emergence of a topological gapless phase. \emph{New Journal of Physics} \textbf{2007}, \emph{9}, 356\relax
\mciteBstWouldAddEndPuncttrue
\mciteSetBstMidEndSepPunct{\mcitedefaultmidpunct}
{\mcitedefaultendpunct}{\mcitedefaultseppunct}\relax
\EndOfBibitem
\bibitem[Xu \latin{et~al.}(2015)Xu, Belopolski, Alidoust, Neupane, Bian, Zhang, Sankar, Chang, Yuan, Lee, Huang, Zheng, Ma, Sanchez, Wang, Bansil, Chou, Shibayev, Lin, Jia, and Hasan]{r15_Xu_WeylF}
Xu,~S.-Y. \latin{et~al.}  Discovery of a Weyl fermion semimetal and topological Fermi arcs. \emph{Science} \textbf{2015}, \emph{349}, 613--617\relax
\mciteBstWouldAddEndPuncttrue
\mciteSetBstMidEndSepPunct{\mcitedefaultmidpunct}
{\mcitedefaultendpunct}{\mcitedefaultseppunct}\relax
\EndOfBibitem
\bibitem[Lv \latin{et~al.}(2015)Lv, Xu, Weng, Ma, Richard, Huang, Zhao, Chen, Matt, Bisti, Strocov, Mesot, Fang, Dai, Qian, Shi, and Ding]{r15_Lv_TaAs}
Lv,~B.~Q. \latin{et~al.}  Observation of Weyl nodes in TaAs. \emph{Nature Physics} \textbf{2015}, \emph{11}, 724--727\relax
\mciteBstWouldAddEndPuncttrue
\mciteSetBstMidEndSepPunct{\mcitedefaultmidpunct}
{\mcitedefaultendpunct}{\mcitedefaultseppunct}\relax
\EndOfBibitem
\bibitem[Xiao \latin{et~al.}(2010)Xiao, Chang, and Niu]{r10_Di_Bphase_rev}
Xiao,~D.; Chang,~M.-C.; Niu,~Q. Berry phase effects on electronic properties. \emph{Rev. Mod. Phys.} \textbf{2010}, \emph{82}, 1959--2007\relax
\mciteBstWouldAddEndPuncttrue
\mciteSetBstMidEndSepPunct{\mcitedefaultmidpunct}
{\mcitedefaultendpunct}{\mcitedefaultseppunct}\relax
\EndOfBibitem
\bibitem[Xie \latin{et~al.}(2021)Xie, Liu, Wang, Cheng, Tian, and Chen]{r21_Xie_TopSemM_rev}
Xie,~B.; Liu,~H.; Wang,~H.; Cheng,~H.; Tian,~J.; Chen,~S. A Review of Topological Semimetal Phases in Photonic Artificial Microstructures. \emph{Frontiers in Physics} \textbf{2021}, \emph{9}\relax
\mciteBstWouldAddEndPuncttrue
\mciteSetBstMidEndSepPunct{\mcitedefaultmidpunct}
{\mcitedefaultendpunct}{\mcitedefaultseppunct}\relax
\EndOfBibitem
\bibitem[Xiao \latin{et~al.}(2020)Xiao, Wang, Wang, Pemmaraju, Wang, Muscher, Sie, Nyby, Devereaux, Qian, Zhang, and Lindenberg]{Xiao:2020gk}
Xiao,~J.; Wang,~Y.; Wang,~H.; Pemmaraju,~C.~D.; Wang,~S.; Muscher,~P.; Sie,~E.~J.; Nyby,~C.~M.; Devereaux,~T.~P.; Qian,~X.; Zhang,~X.; Lindenberg,~A.~M. {Berry curvature memory through electrically driven stacking transitions}. \emph{Nature Physics} \textbf{2020}, \emph{556}, 80--8\relax
\mciteBstWouldAddEndPuncttrue
\mciteSetBstMidEndSepPunct{\mcitedefaultmidpunct}
{\mcitedefaultendpunct}{\mcitedefaultseppunct}\relax
\EndOfBibitem
\bibitem[Ji \latin{et~al.}(2021)Ji, Gr{\aa}n{\"a}s, and Weissenrieder]{Ji:2021cn}
Ji,~S.; Gr{\aa}n{\"a}s,~O.; Weissenrieder,~J. {Manipulation of Stacking Order in Td-WTe2 by Ultrafast Optical Excitation.} \emph{ACS Nano} \textbf{2021}, \emph{15}, 8826--8835\relax
\mciteBstWouldAddEndPuncttrue
\mciteSetBstMidEndSepPunct{\mcitedefaultmidpunct}
{\mcitedefaultendpunct}{\mcitedefaultseppunct}\relax
\EndOfBibitem
\bibitem[Guan \latin{et~al.}(2021)Guan, Wang, You, Sun, and Meng]{ShengMeng:2021}
Guan,~M.-X.; Wang,~E.; You,~P.-W.; Sun,~J.-T.; Meng,~S. Manipulating Weyl quasiparticles by orbital-selective photoexcitation in WTe2. \emph{Nature Communications} \textbf{2021}, \emph{12}, 1885\relax
\mciteBstWouldAddEndPuncttrue
\mciteSetBstMidEndSepPunct{\mcitedefaultmidpunct}
{\mcitedefaultendpunct}{\mcitedefaultseppunct}\relax
\EndOfBibitem
\bibitem[Ji \latin{et~al.}(2022)Ji, Gr{\aa}n{\"a}s, Kumar~Prasad, and Weissenrieder]{ShaozhengJi:2022gb}
Ji,~S.; Gr{\aa}n{\"a}s,~O.; Kumar~Prasad,~A.; Weissenrieder,~J. {Influence of strain on an ultrafast phase transition}. \emph{Nanoscale} \textbf{2022}, \emph{15}, 304--312\relax
\mciteBstWouldAddEndPuncttrue
\mciteSetBstMidEndSepPunct{\mcitedefaultmidpunct}
{\mcitedefaultendpunct}{\mcitedefaultseppunct}\relax
\EndOfBibitem
\bibitem[Gr{\aa}n{\"a}s \latin{et~al.}(2022)Gr{\aa}n{\"a}s, Vaskivskyi, Wang, Thunstr{\"o}m, Ghimire, Knut, S{\"o}derstr{\"o}m, Kjellsson, Turenne, Engel, Beye, Lu, Higley, Reid, Schlotter, Coslovich, Hoffmann, Kolesov, Sch{\"u}{\ss}ler-Langeheine, Styervoyedov, Tancogne-Dejean, Sentef, Reis, Rubio, Parkin, Karis, Rubensson, Eriksson, and D{\"u}rr]{Granas:2022eu}
Gr{\aa}n{\"a}s,~O. \latin{et~al.}  {Ultrafast modification of the electronic structure of a correlated insulator}. \emph{Phys. Rev. Research} \textbf{2022}, \emph{4}, L032030\relax
\mciteBstWouldAddEndPuncttrue
\mciteSetBstMidEndSepPunct{\mcitedefaultmidpunct}
{\mcitedefaultendpunct}{\mcitedefaultseppunct}\relax
\EndOfBibitem
\bibitem[Shin \latin{et~al.}(2019)Shin, Sato, H{\"u}bener, De~Giovannini, Kim, Park, and Rubio]{Shin:2019kt}
Shin,~D.; Sato,~S.~A.; H{\"u}bener,~H.; De~Giovannini,~U.; Kim,~J.; Park,~N.; Rubio,~A. {Unraveling materials Berry curvature and Chern numbers from real-time evolution of Bloch states}. \emph{P Natl Acad Sci Usa} \textbf{2019}, \emph{116}, 4135--4140\relax
\mciteBstWouldAddEndPuncttrue
\mciteSetBstMidEndSepPunct{\mcitedefaultmidpunct}
{\mcitedefaultendpunct}{\mcitedefaultseppunct}\relax
\EndOfBibitem
\bibitem[Tancogne-Dejean \latin{et~al.}(2020)Tancogne-Dejean, Oliveira, Andrade, Appel, Borca, Le~Breton, Buchholz, Castro, Corni, Correa, De~Giovannini, Delgado, Eich, Flick, Gil, Gomez, Helbig, Hübener, Jestädt, Jornet-Somoza, Larsen, Lebedeva, Lüders, Marques, Ohlmann, Pipolo, Rampp, Rozzi, Strubbe, Sato, Schäfer, Theophilou, Welden, and Rubio]{Octopus_code}
Tancogne-Dejean,~N. \latin{et~al.}  {Octopus, a computational framework for exploring light-driven phenomena and quantum dynamics in extended and finite systems}. \emph{The Journal of Chemical Physics} \textbf{2020}, \emph{152}, 124119\relax
\mciteBstWouldAddEndPuncttrue
\mciteSetBstMidEndSepPunct{\mcitedefaultmidpunct}
{\mcitedefaultendpunct}{\mcitedefaultseppunct}\relax
\EndOfBibitem
\bibitem[Sharma \latin{et~al.}(2014)Sharma, Dewhurst, and Gross]{Sharma2014_TDDFT}
Sharma,~S.; Dewhurst,~J.~K.; Gross,~E. K.~U. In \emph{First Principles Approaches to Spectroscopic Properties of Complex Materials}; Di~Valentin,~C., Botti,~S., Cococcioni,~M., Eds.; Springer Berlin Heidelberg: Berlin, Heidelberg, 2014; pp 235--257\relax
\mciteBstWouldAddEndPuncttrue
\mciteSetBstMidEndSepPunct{\mcitedefaultmidpunct}
{\mcitedefaultendpunct}{\mcitedefaultseppunct}\relax
\EndOfBibitem
\bibitem[Onida \latin{et~al.}(2002)Onida, Reining, and Rubio]{r02_Onida_TDDFT_rev}
Onida,~G.; Reining,~L.; Rubio,~A. Electronic excitations: density-functional versus many-body Green's-function approaches. \emph{Rev. Mod. Phys.} \textbf{2002}, \emph{74}, 601--659\relax
\mciteBstWouldAddEndPuncttrue
\mciteSetBstMidEndSepPunct{\mcitedefaultmidpunct}
{\mcitedefaultendpunct}{\mcitedefaultseppunct}\relax
\EndOfBibitem
\bibitem[Kohn(1999)]{r99_Kohn_NLect}
Kohn,~W. Nobel Lecture: Electronic structure of matter---wave functions and density functionals. \emph{Rev. Mod. Phys.} \textbf{1999}, \emph{71}, 1253--1266\relax
\mciteBstWouldAddEndPuncttrue
\mciteSetBstMidEndSepPunct{\mcitedefaultmidpunct}
{\mcitedefaultendpunct}{\mcitedefaultseppunct}\relax
\EndOfBibitem
\bibitem[Runge and Gross(1984)Runge, and Gross]{r84_RungeGross_TDDFT}
Runge,~E.; Gross,~E. K.~U. Density-Functional Theory for Time-Dependent Systems. \emph{Phys. Rev. Lett.} \textbf{1984}, \emph{52}, 997--1000\relax
\mciteBstWouldAddEndPuncttrue
\mciteSetBstMidEndSepPunct{\mcitedefaultmidpunct}
{\mcitedefaultendpunct}{\mcitedefaultseppunct}\relax
\EndOfBibitem
\bibitem[Elk()]{ElkCODE}
The {E}lk {C}ode. \url{https://elk.sourceforge.io}\relax
\mciteBstWouldAddEndPuncttrue
\mciteSetBstMidEndSepPunct{\mcitedefaultmidpunct}
{\mcitedefaultendpunct}{\mcitedefaultseppunct}\relax
\EndOfBibitem
\bibitem[Yu \latin{et~al.}(2011)Yu, Qi, Bernevig, Fang, and Dai]{r11_Yu_z2invariant_berry_connection}
Yu,~R.; Qi,~X.~L.; Bernevig,~A.; Fang,~Z.; Dai,~X. Equivalent expression of ${\mathbb{Z}}_{2}$ topological invariant for band insulators using the non-Abelian Berry connection. \emph{Phys. Rev. B} \textbf{2011}, \emph{84}, 075119\relax
\mciteBstWouldAddEndPuncttrue
\mciteSetBstMidEndSepPunct{\mcitedefaultmidpunct}
{\mcitedefaultendpunct}{\mcitedefaultseppunct}\relax
\EndOfBibitem
\bibitem[Sun \latin{et~al.}(2015)Sun, Wu, and Yan]{r15_Yan_noncentrosymetric_WS}
Sun,~Y.; Wu,~S.-C.; Yan,~B. Topological surface states and Fermi arcs of the noncentrosymmetric Weyl semimetals TaAs, TaP, NbAs, and NbP. \emph{Phys. Rev. B} \textbf{2015}, \emph{92}, 115428\relax
\mciteBstWouldAddEndPuncttrue
\mciteSetBstMidEndSepPunct{\mcitedefaultmidpunct}
{\mcitedefaultendpunct}{\mcitedefaultseppunct}\relax
\EndOfBibitem
\bibitem[Wu \latin{et~al.}(2015)Wu, Ghimire, Reis, Schafer, and Gaarde]{r15_Wu_HoustonSt}
Wu,~M.; Ghimire,~S.; Reis,~D.~A.; Schafer,~K.~J.; Gaarde,~M.~B. High-harmonic generation from Bloch electrons in solids. \emph{Phys. Rev. A} \textbf{2015}, \emph{91}, 043839\relax
\mciteBstWouldAddEndPuncttrue
\mciteSetBstMidEndSepPunct{\mcitedefaultmidpunct}
{\mcitedefaultendpunct}{\mcitedefaultseppunct}\relax
\EndOfBibitem
\bibitem[Krieger and Iafrate(1986)Krieger, and Iafrate]{r86_Krieger_HoustonBasis}
Krieger,~J.~B.; Iafrate,~G.~J. Time evolution of Bloch electrons in a homogeneous electric field. \emph{Phys. Rev. B} \textbf{1986}, \emph{33}, 5494--5500\relax
\mciteBstWouldAddEndPuncttrue
\mciteSetBstMidEndSepPunct{\mcitedefaultmidpunct}
{\mcitedefaultendpunct}{\mcitedefaultseppunct}\relax
\EndOfBibitem
\bibitem[Sie \latin{et~al.}(2019)Sie, Nyby, Pemmaraju, Park, Shen, Yang, Hoffmann, Ofori-Okai, Li, Reid, Weathersby, Mannebach, Finney, Rhodes, Chenet, Antony, Balicas, Hone, Devereaux, Heinz, Wang, and Lindenberg]{r19_Sie_WTe2_WNshifting}
Sie,~E.~J. \latin{et~al.}  An ultrafast symmetry switch in a Weyl semimetal. \emph{Nature} \textbf{2019}, \emph{565}, 61--66\relax
\mciteBstWouldAddEndPuncttrue
\mciteSetBstMidEndSepPunct{\mcitedefaultmidpunct}
{\mcitedefaultendpunct}{\mcitedefaultseppunct}\relax
\EndOfBibitem
\bibitem[Singh and Nordstr\"{o}m(2006)Singh, and Nordstr\"{o}m]{book_SinghNordstrom_LAPW}
Singh,~D.~J.; Nordstr\"{o}m,~L. \emph{Planewaves, Pseudopotentials, and the LAPW method}; Springer Science \& Business Media, 2006\relax
\mciteBstWouldAddEndPuncttrue
\mciteSetBstMidEndSepPunct{\mcitedefaultmidpunct}
{\mcitedefaultendpunct}{\mcitedefaultseppunct}\relax
\EndOfBibitem
\bibitem[Elliott \latin{et~al.}(2016)Elliott, M{\"u}ller, Dewhurst, Sharma, and Gross]{Elliott2016_TDDFT}
Elliott,~P.; M{\"u}ller,~T.; Dewhurst,~J.~K.; Sharma,~S.; Gross,~E. K.~U. Ultrafast laser induced local magnetization dynamics in Heusler compounds. \emph{Scientific Reports} \textbf{2016}, \emph{6}, 38911\relax
\mciteBstWouldAddEndPuncttrue
\mciteSetBstMidEndSepPunct{\mcitedefaultmidpunct}
{\mcitedefaultendpunct}{\mcitedefaultseppunct}\relax
\EndOfBibitem
\bibitem[Dewhurst \latin{et~al.}(2021)Dewhurst, Shallcross, Elliott, Eisebitt, Schmising, and Sharma]{Dewhurst2021_magdyn}
Dewhurst,~J.~K.; Shallcross,~S.; Elliott,~P.; Eisebitt,~S.; Schmising,~C. v.~K.; Sharma,~S. Angular momentum redistribution in laser-induced demagnetization. \emph{Phys. Rev. B} \textbf{2021}, \emph{104}, 054438\relax
\mciteBstWouldAddEndPuncttrue
\mciteSetBstMidEndSepPunct{\mcitedefaultmidpunct}
{\mcitedefaultendpunct}{\mcitedefaultseppunct}\relax
\EndOfBibitem
\bibitem[Perdew \latin{et~al.}(1996)Perdew, Burke, and Ernzerhof]{r96_Perdew_PBE}
Perdew,~J.~P.; Burke,~K.; Ernzerhof,~M. Generalized Gradient Approximation Made Simple. \emph{Phys. Rev. Lett.} \textbf{1996}, \emph{77}, 3865--3868\relax
\mciteBstWouldAddEndPuncttrue
\mciteSetBstMidEndSepPunct{\mcitedefaultmidpunct}
{\mcitedefaultendpunct}{\mcitedefaultseppunct}\relax
\EndOfBibitem
\bibitem[w_S()]{w_Spinger_crystdata}
TaAs Crystal Structure: Datasheet from ``PAULING FILE Multinaries Edition -- 2022'' in SpringerMaterials (https://materials.springer.com/isp/crystallographic/docs/sd{\_}0452697). \url{https://materials.springer.com/isp/crystallographic/docs/sd_0452697}, Copyright 2016 Springer-Verlag Berlin Heidelberg {\&} Material Phases Data System (MPDS), Switzerland {\&} National Institute for Materials Science (NIMS), Japan\relax
\mciteBstWouldAddEndPuncttrue
\mciteSetBstMidEndSepPunct{\mcitedefaultmidpunct}
{\mcitedefaultendpunct}{\mcitedefaultseppunct}\relax
\EndOfBibitem
\bibitem[Furuseth \latin{et~al.}(1965)Furuseth, Selte, and Kjekshus]{r65_Furuseth_TaAs_cryst}
Furuseth,~S.; Selte,~K.; Kjekshus,~A. ON THE ARSENIDES AND ANTIMONIDES OF TANTALUM. \emph{Acta Chemica Scandinavica (Denmark) Divided into Acta Chem. Scand., Ser. A and Ser. B} \textbf{1965}, \emph{Vol: 19}\relax
\mciteBstWouldAddEndPuncttrue
\mciteSetBstMidEndSepPunct{\mcitedefaultmidpunct}
{\mcitedefaultendpunct}{\mcitedefaultseppunct}\relax
\EndOfBibitem
\bibitem[Baer and Kronik(2018)Baer, and Kronik]{r18_Baer_Tddft}
Baer,~R.; Kronik,~L. Time-dependent generalized Kohn--Sham theory. \emph{The European Physical Journal B} \textbf{2018}, \emph{91}, 170\relax
\mciteBstWouldAddEndPuncttrue
\mciteSetBstMidEndSepPunct{\mcitedefaultmidpunct}
{\mcitedefaultendpunct}{\mcitedefaultseppunct}\relax
\EndOfBibitem
\bibitem[Kohn and Sham(1965)Kohn, and Sham]{r65_KohnSham}
Kohn,~W.; Sham,~L.~J. Self-Consistent Equations Including Exchange and Correlation Effects. \emph{Phys. Rev.} \textbf{1965}, \emph{140}, A1133--A1138\relax
\mciteBstWouldAddEndPuncttrue
\mciteSetBstMidEndSepPunct{\mcitedefaultmidpunct}
{\mcitedefaultendpunct}{\mcitedefaultseppunct}\relax
\EndOfBibitem
\bibitem[Dewhurst \latin{et~al.}(2016)Dewhurst, Krieger, Sharma, and Gross]{r16_Dewhurst_ElkTimeEvol}
Dewhurst,~J.; Krieger,~K.; Sharma,~S.; Gross,~E. An efficient algorithm for time propagation as applied to linearized augmented plane wave method. \emph{Computer Physics Communications} \textbf{2016}, \emph{209}, 92--95\relax
\mciteBstWouldAddEndPuncttrue
\mciteSetBstMidEndSepPunct{\mcitedefaultmidpunct}
{\mcitedefaultendpunct}{\mcitedefaultseppunct}\relax
\EndOfBibitem
\bibitem[F\o{}rre and Simonsen(2016)F\o{}rre, and Simonsen]{r16_Forre_VelG}
F\o{}rre,~M.; Simonsen,~A.~S. Generalized velocity-gauge form of the light-matter interaction Hamiltonian beyond the dipole approximation. \emph{Phys. Rev. A} \textbf{2016}, \emph{93}, 013423\relax
\mciteBstWouldAddEndPuncttrue
\mciteSetBstMidEndSepPunct{\mcitedefaultmidpunct}
{\mcitedefaultendpunct}{\mcitedefaultseppunct}\relax
\EndOfBibitem
\bibitem[Mattiat and Luber(2022)Mattiat, and Luber]{r22_Mattiat_VelG}
Mattiat,~J.; Luber,~S. Comparison of Length, Velocity, and Symmetric Gauges for the Calculation of Absorption and Electric Circular Dichroism Spectra with Real-Time Time-Dependent Density Functional Theory. \emph{Journal of Chemical Theory and Computation} \textbf{2022}, \emph{18}, 5513--5526\relax
\mciteBstWouldAddEndPuncttrue
\mciteSetBstMidEndSepPunct{\mcitedefaultmidpunct}
{\mcitedefaultendpunct}{\mcitedefaultseppunct}\relax
\EndOfBibitem
\bibitem[Yu \latin{et~al.}(1991)Yu, Singh, and Krakauer]{r91_Yu_LAPW_forces}
Yu,~R.; Singh,~D.; Krakauer,~H. All-electron and pseudopotential force calculations using the linearized-augmented-plane-wave method. \emph{Phys. Rev. B} \textbf{1991}, \emph{43}, 6411--6422\relax
\mciteBstWouldAddEndPuncttrue
\mciteSetBstMidEndSepPunct{\mcitedefaultmidpunct}
{\mcitedefaultendpunct}{\mcitedefaultseppunct}\relax
\EndOfBibitem
\bibitem[Sharma \latin{et~al.}(2022)Sharma, Shallcross, Elliott, and Dewhurst]{r21_Sharma_tddft_FePt}
Sharma,~S.; Shallcross,~S.; Elliott,~P.; Dewhurst,~J.~K. Making a case for femto-phono-magnetism with FePt. \emph{Science Advances} \textbf{2022}, \emph{8}, eabq2021\relax
\mciteBstWouldAddEndPuncttrue
\mciteSetBstMidEndSepPunct{\mcitedefaultmidpunct}
{\mcitedefaultendpunct}{\mcitedefaultseppunct}\relax
\EndOfBibitem
\bibitem[Fukui \latin{et~al.}(2005)Fukui, Hatsugai, and Suzuki]{r05_Takahiro_DiscrChN}
Fukui,~T.; Hatsugai,~Y.; Suzuki,~H. Chern Numbers in Discretized Brillouin Zone: Efficient Method of Computing (Spin) Hall Conductances. \emph{Journal of the Physical Society of Japan} \textbf{2005}, \emph{74}, 1674--1677\relax
\mciteBstWouldAddEndPuncttrue
\mciteSetBstMidEndSepPunct{\mcitedefaultmidpunct}
{\mcitedefaultendpunct}{\mcitedefaultseppunct}\relax
\EndOfBibitem
\bibitem[Sheng \latin{et~al.}(2006)Sheng, Weng, Sheng, and Haldane]{r06_Sheng_ChN}
Sheng,~D.~N.; Weng,~Z.~Y.; Sheng,~L.; Haldane,~F. D.~M. Quantum Spin-Hall Effect and Topologically Invariant Chern Numbers. \emph{Phys. Rev. Lett.} \textbf{2006}, \emph{97}, 036808\relax
\mciteBstWouldAddEndPuncttrue
\mciteSetBstMidEndSepPunct{\mcitedefaultmidpunct}
{\mcitedefaultendpunct}{\mcitedefaultseppunct}\relax
\EndOfBibitem
\bibitem[Wang \latin{et~al.}(2020)Wang, Zhang, Yuan, Zhong, and Lu]{r20_Wang_ChN}
Wang,~C.; Zhang,~H.; Yuan,~H.; Zhong,~J.; Lu,~C. Universal numerical calculation method for the Berry curvature and Chern numbers of typical topological photonic crystals. \emph{Frontiers of Optoelectronics} \textbf{2020}, \emph{13}, 73--88\relax
\mciteBstWouldAddEndPuncttrue
\mciteSetBstMidEndSepPunct{\mcitedefaultmidpunct}
{\mcitedefaultendpunct}{\mcitedefaultseppunct}\relax
\EndOfBibitem
\bibitem[Zhao \latin{et~al.}(2020)Zhao, Xie, Chen, Lan, Huang, and Sha]{r20_Zhao_ChN}
Zhao,~R.; Xie,~G.-D.; Chen,~M. L.~N.; Lan,~Z.; Huang,~Z.; Sha,~W. E.~I. First-principle calculation of Chern number in gyrotropic photonic crystals. \emph{Opt. Express} \textbf{2020}, \emph{28}, 4638--4649\relax
\mciteBstWouldAddEndPuncttrue
\mciteSetBstMidEndSepPunct{\mcitedefaultmidpunct}
{\mcitedefaultendpunct}{\mcitedefaultseppunct}\relax
\EndOfBibitem
\bibitem[Bradlyn and Iraola(2022)Bradlyn, and Iraola]{r22_Bradlyn_PhysLectNotes}
Bradlyn,~B.; Iraola,~M. {Lecture notes on Berry phases and topology}. \emph{SciPost Phys. Lect. Notes} \textbf{2022}, 51\relax
\mciteBstWouldAddEndPuncttrue
\mciteSetBstMidEndSepPunct{\mcitedefaultmidpunct}
{\mcitedefaultendpunct}{\mcitedefaultseppunct}\relax
\EndOfBibitem
\bibitem[Ivanov and Savrasov(2019)Ivanov, and Savrasov]{r19_Ivanov_WPminimg}
Ivanov,~V.; Savrasov,~S.~Y. Monopole mining method for high-throughput screening for Weyl semimetals. \emph{Phys. Rev. B} \textbf{2019}, \emph{99}, 125124\relax
\mciteBstWouldAddEndPuncttrue
\mciteSetBstMidEndSepPunct{\mcitedefaultmidpunct}
{\mcitedefaultendpunct}{\mcitedefaultseppunct}\relax
\EndOfBibitem
\bibitem[Gresch \latin{et~al.}(2017)Gresch, Aut\`es, Yazyev, Troyer, Vanderbilt, Bernevig, and Soluyanov]{r17_Gresch_Z2pack}
Gresch,~D.; Aut\`es,~G.; Yazyev,~O.~V.; Troyer,~M.; Vanderbilt,~D.; Bernevig,~B.~A.; Soluyanov,~A.~A. Z2Pack: Numerical implementation of hybrid Wannier centers for identifying topological materials. \emph{Phys. Rev. B} \textbf{2017}, \emph{95}, 075146\relax
\mciteBstWouldAddEndPuncttrue
\mciteSetBstMidEndSepPunct{\mcitedefaultmidpunct}
{\mcitedefaultendpunct}{\mcitedefaultseppunct}\relax
\EndOfBibitem
\bibitem[Petersilka \latin{et~al.}(1996)Petersilka, Gossmann, and Gross]{r96_Petersilka_response}
Petersilka,~M.; Gossmann,~U.~J.; Gross,~E. K.~U. Excitation Energies from Time-Dependent Density-Functional Theory. \emph{Phys. Rev. Lett.} \textbf{1996}, \emph{76}, 1212--1215\relax
\mciteBstWouldAddEndPuncttrue
\mciteSetBstMidEndSepPunct{\mcitedefaultmidpunct}
{\mcitedefaultendpunct}{\mcitedefaultseppunct}\relax
\EndOfBibitem
\bibitem[Dewhurst \latin{et~al.}(2020)Dewhurst, Willems, Elliott, Li, Schmising, Str\"uber, Engel, Eisebitt, and Sharma]{r20_Dewhurst_MagDischro}
Dewhurst,~J.~K.; Willems,~F.; Elliott,~P.; Li,~Q.~Z.; Schmising,~C. v.~K.; Str\"uber,~C.; Engel,~D.~W.; Eisebitt,~S.; Sharma,~S. Element Specificity of Transient Extreme Ultraviolet Magnetic Dichroism. \emph{Phys. Rev. Lett.} \textbf{2020}, \emph{124}, 077203\relax
\mciteBstWouldAddEndPuncttrue
\mciteSetBstMidEndSepPunct{\mcitedefaultmidpunct}
{\mcitedefaultendpunct}{\mcitedefaultseppunct}\relax
\EndOfBibitem
\end{mcitethebibliography}
